\RequirePackage[l2tabu, orthodox]{nag}
\documentclass{stsci_report}

\usepackage{booktabs}
\usepackage{rotating}
\usepackage[round, sort]{natbib}
\usepackage{url}
\usepackage{graphicx}
\usepackage[font=footnotesize,labelfont=bf]{caption}
\usepackage{subcaption}
\usepackage{nameref}
\usepackage{rotating}
\usepackage[pdftex, pdfauthor={J. E. Ryon, N. A. Grogin}, pdftitle={Serial Charge Transfer Efficiency in ACS/WFC}, pdfsubject={We present a dedicated study of CCD serial ($x$-direction) charge transfer efficiency (CTE) in ACS/WFC. Following past studies of parallel ($y$-direction) CTE, we use the serial CTE trails behind hot pixels in calibration dark frames to characterize charge trapping and release in the serial registers of the WFC detectors. Serial CTE trails are sharper and longer than parallel CTE trails. Many fewer charge traps come into play during serial pixel transfers than parallel transfers, which explains why parallel CTE is much worse than serial CTE. We find that serial CTE can cause losses of $\sim$0.005-0.02~mag in stellar photometry and shift stellar centroids by $\sim$0.01-0.035 pixels. The pixel-based algorithm in CALACS that corrects for parallel CTE losses in WFC data has been modified to include a correction for serial CTE losses. The PCTETAB reference file has also been updated to include serial CTE parameters. The pixel-based correction for serial CTE currently runs only on full-frame WFC images obtained after SM4 (May 2009). Shortly following the publication of this report, science data corrected for both parallel and serial CTE will be available in the MAST archive.}, pdfkeywords={ACS, WFC, CCD, charge transfer efficiency, serial}]{hyperref} 
\usepackage{array} 
\usepackage{color}
\usepackage{amsmath, amssymb}
\usepackage{aas_macros}
\usepackage{multirow}

\copyrighttext{Copyright \copyright \the\year\ The Association of Universities for Research in Astronomy, Inc. All Rights Reserved.}

\title{\textbf{Serial Charge Transfer Efficiency in ACS/WFC}}
\presubtitle{\flushleft{\includegraphics[width=5cm]{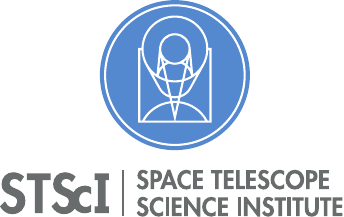}} \\ \hfill Instrument Science Report ACS 2024-07}
\author{J. E. Ryon \& N. A. Grogin}
\date{\today}

\begin{document}

\maketitle

\abstract{We present a dedicated study of CCD serial ($x$-direction) charge transfer efficiency (CTE) in ACS/WFC. Following past studies of parallel ($y$-direction) CTE, we use the serial CTE trails behind hot pixels in calibration dark frames to characterize charge trapping and release in the serial registers of the WFC detectors. Serial CTE trails are sharper and longer than parallel CTE trails. Many fewer charge traps come into play during serial pixel transfers than parallel transfers, which explains why parallel CTE is much worse than serial CTE. We find that serial CTE can cause losses of $\sim$0.005-0.02~mag in stellar photometry and shift stellar centroids by $\sim$0.01-0.035 pixels. The pixel-based algorithm in CALACS that corrects for parallel CTE losses in WFC data has been modified to include a correction for serial CTE losses. The PCTETAB reference file has also been updated to include serial CTE parameters. The pixel-based correction for serial CTE currently runs only on full-frame WFC images obtained after SM4 (May 2009). Shortly following the publication of this report, science data corrected for both parallel and serial CTE will be available in the MAST archive.}

\section{Introduction} \label{intro}

Charge transfer efficiency (CTE) in the Advanced Camera for Surveys (ACS) Wide Field Channel (WFC) has been extensively studied over the instrument's lifetime. The primary focus of past studies was to characterize and correct imperfect CTE in the CCD parallel transfer direction ($y$-direction). The period of time between parallel pixel transfers (row to row), also called ``dwell time", is long compared to charge trapping and release timescales, which leads to bright trails of charge extending from most sources along the columns of the detector. Photometry and astrometry of sources are significantly affected, especially faint sources on low backgrounds \citep{anderson2010, anderson2018, stark2024}. 

The dwell time between serial transfers ($x$-direction) is very short, so fewer electrons are trapped and released than during parallel transfers. The effects of imperfect serial CTE are therefore relatively minor compared to parallel CTE, but as CTE in both transfer directions degrades with time, a dedicated study of serial CTE is now warranted.

This work relies heavily on prior work by \cite{anderson2010} and \cite{anderson2018} on the parallel CTE model. We follow a similar procedure as laid out in \cite{anderson2018} for determining the charge trap density and trail profiles for an empirical serial CTE model. We also extend the existing software in CALACS for pixel-based parallel CTE correction, ACSCTE, to correct serial CTE losses in ACS/WFC data.

In Section~\ref{background}, we provide background on the ACS/WFC readout process and the detector effects that cause CTE losses. We describe the empirical CTE model and pixel-based correction algorithm in Section~\ref{model}. The data used to determine the serial CTE model parameters are described in Section~\ref{data}. The analysis of these data and the resulting model parameters are detailed in Section~\ref{analysis}. In Section~\ref{results}, we discuss the qualitative and quantitative impact of serial CTE correction on data. We further describe the implementation of serial CTE correction in CALACS in Section~\ref{implementation} and conclude in Section~\ref{conclusions}. 

\section{Background} \label{background}

\subsection{ACS/WFC Full-frame Readout} \label{readout}

\begin{figure}[t]
  \centering
  \includegraphics[width=0.7\textwidth]{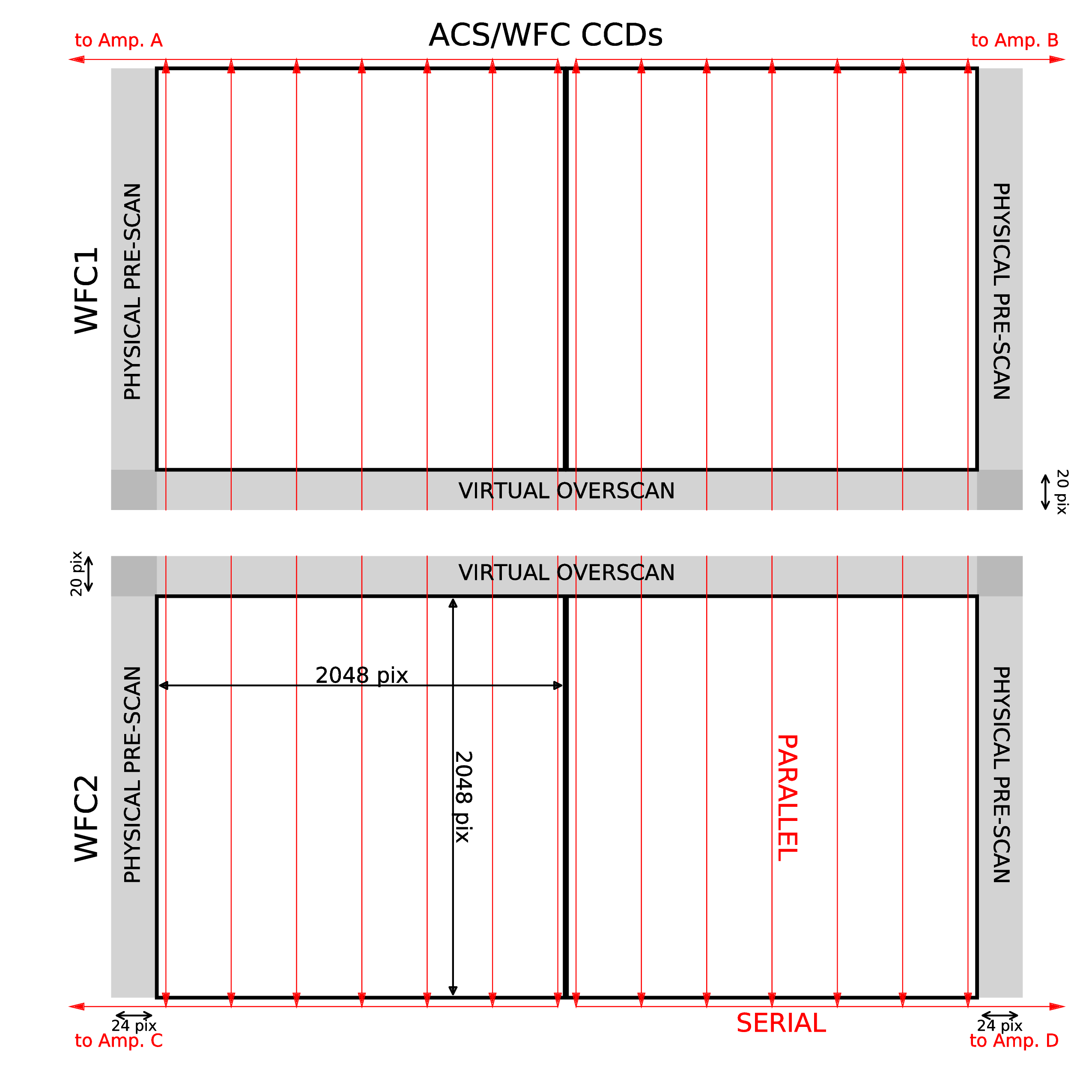}
    \caption{Schematic of the ACS/WFC CCDs including active pixel arrays (white regions), prescans and overscans (gray regions), and parallel and serial transfer directions (red arrows). Reproduced from the ACS Instrument Handbook \citep{stark2024b}.}
    \label{wfc_layout}
\end{figure}

As shown in Figure~\ref{wfc_layout}, WFC consists of two charge-coupled devices (CCDs), each containing 4096$\times$2048 pixels. Each CCD also has an associated serial readout register consisting of 4144 additional pixels (including 24 prescan pixels at either end). The serial registers are not exposed to the sky during an exposure, but they hold and transfer charge in much the same way as the active, on-sky pixels. The WFC1 serial register is located at the top of the WFC full-frame, and the WFC2 serial register at the bottom. 

Two amplifiers read out each CCD: amplifiers A and B each read out half of the WFC1 array, and amplifiers C and D do the same for WFC2. (We will refer to the quadrants of the active pixel arrays by their amplifier letter in the following.) When an exposure is complete, the first parallel transfer of charge of the active pixel arrays takes place. Charge in all rows is transferred one row towards the serial registers vertically, or in the $y$-direction, with the first row's charge being deposited into the serial register. The charge in the serial register is then transferred to the amplifiers serially (in the horizontal, or $x$-direction), one pixel at a time. Once the first row of pixels is read out, the second parallel transfer of the entire array takes place, followed by serial transfers of the second row to the amplifiers, and so on for the rest of the array. The pixel transfer directions for both chips are shown by red arrows in Figure~\ref{wfc_layout}.

\subsection{Charge Transfer Efficiency} \label{cte_background}

If charge transfer were perfectly efficient, all charge accumulated during an exposure would reach the amplifiers and be located within the pixels in which it was originally deposited. However, as the CCDs are damaged over time by particle impacts from HST's low-Earth orbit radiation environment, CTE degrades. Damage to the CCDs gives rise to charge traps within detector pixels (and enhanced charge generation, i.e., warm and hot pixels).

Charge traps capture electrons from the packets of charge within each pixel as they are transferred through pixels during readout. As described in \cite{massey2010}, packets of charge containing more electrons are physically larger than charge packets with fewer electrons. Traps can be located throughout the 3D space within a pixel. Traps near the center of a pixel are very likely to capture electrons from any size charge packet, whereas traps on the outskirts of a pixel may only trap electrons from the largest of charge packets. The number of traps affecting each size of charge packet, called the trap density profile, is one key feature of the empirical CTE model described in \cite{anderson2010} and \cite{anderson2018}.

\begin{figure}[t]
  \centering
  \includegraphics[width=0.8\textwidth]{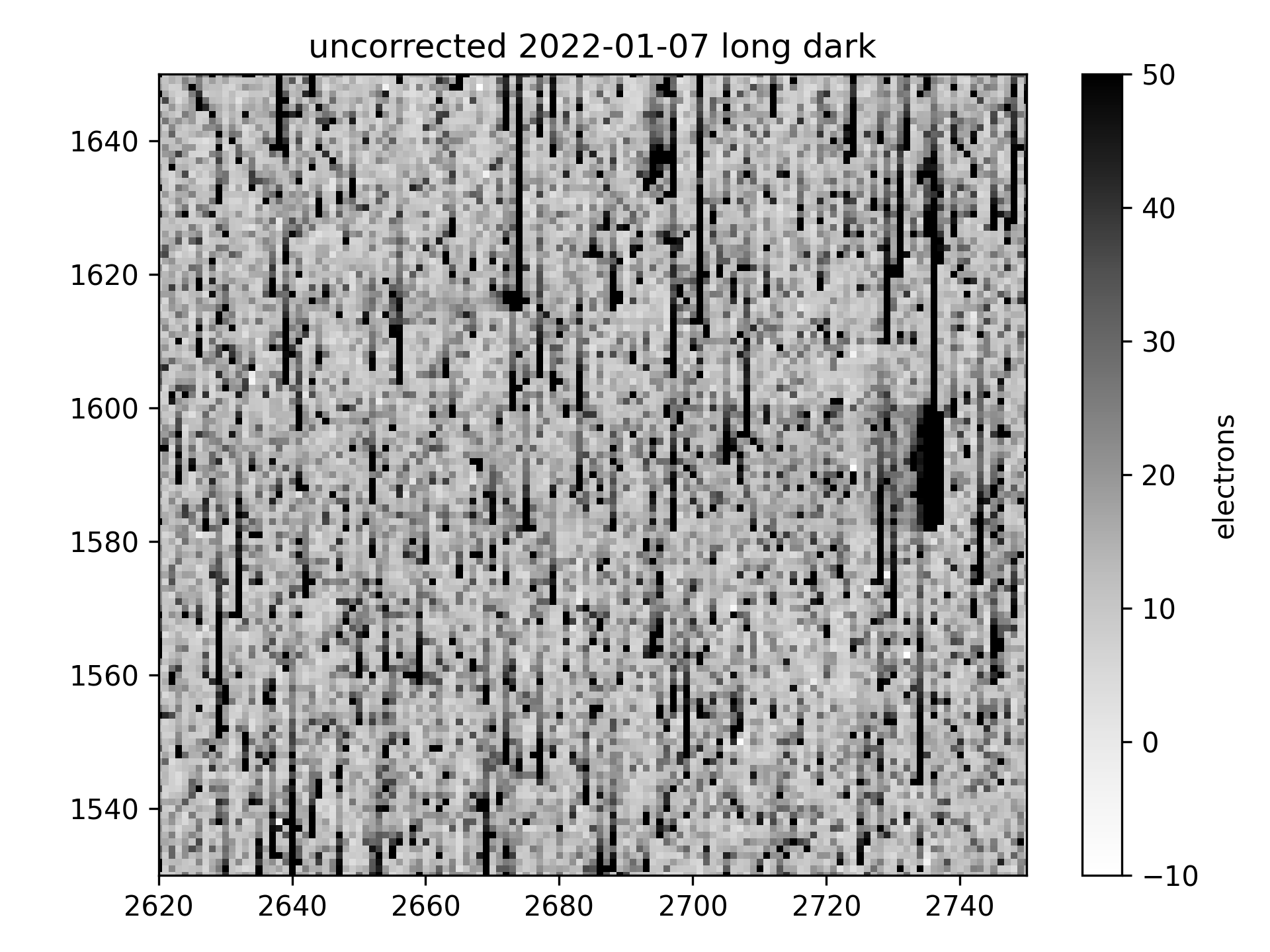}
    \caption{Section of quadrant D in a 2022 stack of 48 long dark frames (1000.5s each) corrected for post-flash. Hot pixels and CTE trails are dark gray and black in this grayscale image. Parallel and serial CTE trails extend towards the top and left of the image, respectively. Parallel trails are visible above the background for several pixels, often 20 or more. Serial trails are typically evident as a single bright pixel immediately to the left of hot pixels. A notable exception is the serial trail of the highly saturated hot pixel located at approximately (2730, 1590), which is faintly visible for almost 40 pixels.}
    \label{trail_image}
\end{figure}

Trapped electrons are released some time later, into subsequent pixels of the original scene. This charge capture and delayed release causes trails of charge to emanate from sources in the image, in the direction opposite to the parallel and serial transfer directions, as shown in Figure~\ref{trail_image}. CTE trails affect photometry and astrometry of astronomical sources by reducing the flux associated with them and shifting their centroids. Determining the probability of charge release as a function of pixel location relative to a charge packet, called the trail profile, is another key feature of the empirical CTE model.

Charge traps in the serial registers cause serial CTE losses. Because all charge accumulated in the detector during an exposure will travel through the serial register during readout, there is a large amount of data with which to characterize serial CTE, though it is a much weaker effect than parallel CTE. In this work, we determine the trap density and trail profiles appropriate for each quadrant's serial register separately, since each has a distinct history of radiation damage.

CTE is dependent on many factors, including temperature of the detectors, source signal, background level, and dwell time between pixel transfers during readout. The dwell time is particularly relevant for the difference between parallel and serial CTE because it determines the time available for charge trapping and release to occur in a given pixel. During typical operation of ACS/WFC, one parallel transfer takes 3212~$\mu$s, while one serial transfer takes 22~$\mu$s \citep{mutchler2005}. The time constants for charge trapping for known species of traps are $\leq$10~$\mu$s for charge packets $\geq$50~e$^-$ in size at the WFC operating temperature of $-$80$^\circ$C \citep{cawley2001}. The time constants for smaller charge packets, $<50$~e$^-$, however, are comparable to or somewhat longer than the serial transfer dwell time ($\sim$60~$\mu$s). This means that fewer traps should come into play during serial transfers than parallel, especially for small charge packets. Charge release timescales appear to vary significantly for known trap species at $-80^\circ$C \citep{cawley2001}, but some are much longer than the serial dwell time \citep{massey2010b}, so serial CTE trails may be longer than parallel. 

Pixels with anomalously high dark current, called warm and hot pixels, are very useful for probing charge trapping and release. Hot pixels are essentially delta functions of charge, unlike stars, and span a wide range of amplitudes. Studying the total deferred charge in hot pixel CTE trails as a function of hot pixel brightness is the basis for characterizing serial CTE, as further discussed in Section~\ref{analysis}.

Serial CTE in ACS/WFC was briefly studied in \cite{anderson2010}. At that time, the first pixel in serial CTE trails following hot pixels contained about 0.4 - 1\% of the signal of the hot pixel, and subsequent pixels in the trail contained very little detectable signal. It was concluded that the effect on photometry is negligible, and the effect on astrometry is small but non-negligible, though neither effect was quantified. Extending the model to include serial CTE was left to future work. In the latest update to the photometric CTE correction model for point sources, serial CTE losses were noted as being an order of magnitude smaller than parallel CTE losses and therefore difficult to accurately measure \citep{chiaberge2022}. Another recent study focused on modeling photometric repeatability of ACS/WFC observations of 47 Tuc found strong evidence for serial CTE losses in recent data, but the amplitude of the effect was not able to be precisely quantified \citep{ryan2024}.

\section{CTE Model} \label{model}

The current pixel-based CTE model implemented in CALACS to correct for parallel CTE losses is described in detail in \citep{anderson2018} and the original version in \cite{anderson2010}. Here, we briefly describe the key features of the model and changes for serial CTE. 

ACSCTE is the step within CALACS that applies the pixel-based model for CTE correction to data. ACSCTE runs after ACSCCD if the PCTECORR header keyword is set to PERFORM. ACSCCD initializes the ERR and DQ arrays, performs bias correction, and applies the amplifier gain correction to convert from data numbers (DN) to electrons. The resulting BLV\_TMP files are input into ACSCTE along with the PCTETAB reference file, which contains the necessary parameters for the CTE correction model. The PCTETAB includes the trap density profile, trail profile, constants for the time-dependent scaling of the model (CTEDATE0/1), number of forward-model iterations (PCTENFOR), and number of parallel transfer stages (PCTENPAR). 

The algorithm implemented in ACSCTE is an iterative forward model of the readout process. For a given input BLV\_TMP image $I$, the forward model applies further CTE trailing, resulting in image $R$. The difference image, $D = R - I$, is subtracted from the input to produce a corrected image, $C = I - D$. This process is iterated according to the number of forward model iterations specified in the header keyword PCTENFOR (currently five for the parallel CTE model). For the serial CTE model, we perform a single iteration, since the serial CTE losses are significantly smaller than parallel.

To simulate the readout process, the pixel-based CTE model essentially tracks the state of each charge trap as charge packets are transferred through pixels. If a charge packet is large enough to encounter a given charge trap, and the trap is empty (i.e., it hasn't encountered a packet that size yet), the trap will capture an electron from the packet. The electron is released into a subsequent pixel as readout continues, if the subsequent pixel contains a smaller charge packet than the trap affects.

In reality, the locations of individual charge traps in the serial registers are unknown, so all pixels in each register must be treated as having the same distribution of traps. Detailed simulating of every trap in every pixel for all CCD transfers far exceeds current computational capacity. In practice, traps are approximated to capture fractional electrons and release fractional electrons according to the probability of release in the trail profile. The algorithm is further simplified by scaling the ``intensity'' of trapping according to the column number, i.e., instead of tracking a charge packet at $x=100$ through 100 transfers to the register, the packet encounters traps with 100$\times$ the ``strength'' of traps at $x=1$. 

For the parallel model, charge packets located far from the amplifiers, especially smaller ones, will lose so many electrons that the charge trap distribution encountered by the packet will change significantly as the packet is transferred along. For this reason, the parallel transfers are broken up into seven stages, according to the PCTENPAR header keyword. For the serial CTE model, we set PCTENPAR to one, since the serial CTE losses are significantly smaller than parallel.

CTE losses in the parallel direction increase linearly with time \citep{ryon2017, chiaberge2009, mutchler2005}, so we expect the same for serial CTE. Time dependence is included by linearly scaling the intensity of the traps according to the factor $s$,
\begin{equation}
s = (t_{\mathrm{obs}} - t_0)/(t_1 - t_0),
\end{equation}
where $t_{\mathrm{obs}}$ is the MJD of the observation. For parallel CTE, $t_0$ and $t_1$ correspond to the MJD of ACS installation on HST (MJD 52334.8) and the observation MJD of the data from which the model parameters were determined (the ``pinning'' date, MJD 57710.4), respectively. For serial CTE, we adjust $t_0$ and $t_1$ manually in order to fully address the time dependence, as discussed further in Section~\ref{time_dependence}. These constants are listed in the header keywords CTEDATE0 and CTEDATE1 in the PCTETAB.

To extend the ACSCTE algorithm to correct serial CTE losses, each quadrant of the image is rotated by 90 degrees, such that the rows become columns and the serial transfer direction becomes the parallel transfer direction. The serial correction step is run, the serial-corrected quadrants are rotated back to their native orientation, and then the parallel correction step is run. Parallel pixel transfers precede serial pixel transfers during readout, so serial CTE losses can affect charge deferred during parallel transfers. For this reason, to truly reverse the readout process, serial correction takes place prior to parallel correction. This ensures that the parallel correction runs on  We further discuss the implementation of serial CTE correction in CALACS in Section~\ref{implementation}.

\section{Data} \label{data}

We use hot pixels in dark exposures from the CCD Daily Monitor calibration program to study serial CTE. In the month-long period between anneals, about 48 darks are taken for the purpose of creating a reference file for dark correction of WFC science data. Prior to January 2015, the exposure time of these long darks was 1040s. Since January 2015, the exposure time is 1000.5s. For the 1000.5s darks, a post-flash background of $\sim$55~e$^-$ is added to fill low-level charge traps to improve CTE, and therefore improve dark rate characterization. For all anneal periods, the long darks are combined into a single image with the ACS/WFC reference file pipeline \citep{desjardins2018}.

Since January 2015, about 48 short (0.5s) darks, post-flashed to the same $\sim$55~e$^-$ level, are also taken during each anneal period. The short darks are also combined into a single image, and the short dark stack is subtracted from the long dark stack to precisely remove the post-flash.

For all anneals, the long dark stack (flash removed if necessary) is normalized by the total exposure time and hot, warm, and unstable pixels are flagged in the data quality (DQ) extensions. The resulting image is the dark reference file for dark correction of science data.

Isolated hot and warm pixels for this study are selected from the dark reference file (non-CTE-corrected DRK, see Section~\ref{hot_pixels}). Once the hot pixels are selected, we study their serial CTE trails in the long dark stack, flash-corrected with the FLSHCORR step in CALACS if necessary (see Section~\ref{trail} and Figure~\ref{trail_image}).

\section{Analysis} \label{analysis}

\subsection{Selecting Hot Pixels} \label{hot_pixels}

We create a mask from the DQ arrays of the dark reference file that includes hot and warm pixels (DQ values 16 or 64) and excludes unstable pixels (DQ value 32). Separately, we apply a filter to each pixel in the SCI arrays that checks the neighboring pixels within $\pm$3 pixels in the $x$-direction and $\pm$5 pixels in the $y$-direction. If all of these neighboring pixels are $<$25\% of the central pixel's signal level, then it is deemed ``isolated'', and added to an isolated pixel mask. The DQ mask and the isolated pixel mask are combined to ensure that only isolated, stable, hot and warm pixels remain. A catalog of $xy$ coordinates, WFC chip, and number of electrons in the pixel is recorded. Catalogs are created for the dark reference file from each anneal.

\subsection{Measuring Serial CTE Trails} \label{trail}

\begin{figure}[t]
  \centering
  \includegraphics[width=0.7\textwidth]{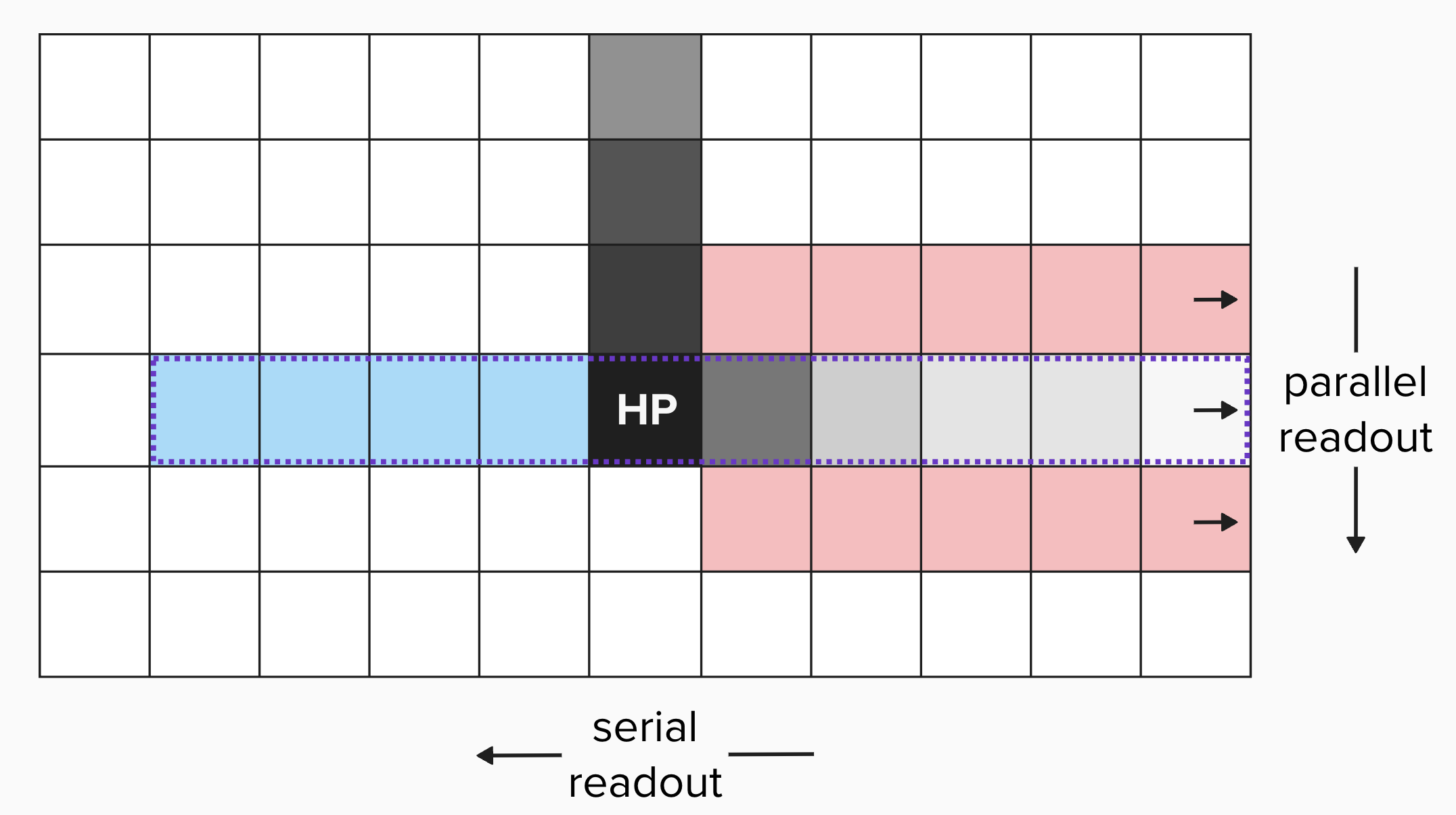}
    \caption{Schematic of a pixel grid including a hot pixel (black pixel labeled HP) and serial and parallel CTE trails. The serial and parallel readout directions are to the left and down, respectively, as shown by the arrows. The serial and parallel trails (grayscale pixels) extend to the right and top of the panel, respectively. Pixels selected for trail analysis are outlined with a dashed purple line, including the four ``upstream'' pixels (light blue), the hot pixel, and 100 pixels ``downstream'', i.e., the serial trail. 100 background pixels (light red) are selected from the rows above and below the serial trail.}
    \label{extract}
\end{figure}

Because the serial trails are quite faint, we combine data from an entire calendar year to detect them. We begin with the long dark stack (flash-corrected, if applicable). From the hot pixel catalog, we select very hot pixels, 30,000$\pm$6,000~e$^-$ (within 20\%), that experience 1500-1900 serial transfers. These should have relatively bright serial trails since they are transferred through most of the serial register. For each pixel, we extract the hot pixel itself, four pixels ``upstream'' of the hot pixel (towards the amplifier), and 100 pixels ``downstream'' of the hot pixel (away from the amplifier) from the long dark. We also extract the rows above and below ($\pm$1 pixel in $y$) the hot pixel's serial trail, avoiding the parallel trail, for background estimation. See Figure~\ref{extract} for a schematic of the selected pixels.

Taking extracted trails from the long dark stacks for a given calendar year, we sigma-clip (3$\sigma$) the pixel values at each location in the trail, then run kernel density estimation (KDE) on the remaining values. KDE is a non-parametric method for estimating the underlying distribution of pixel values, and allows us to estimate the peak of the distribution more robustly than simply choosing the peak of the histogram (which depends strongly on bin width). Additionally, the mean and median are biased toward the long, high-valued tail. We use \texttt{FFTKDE} within the python package \texttt{KDEpy}, and assume a gaussian kernel and Scott's rule of thumb for the kernel bandwidth. The same KDE peak-finding routine is run on the background regions to find a best background estimate.

\begin{figure}[t]
  \centering
  \includegraphics[width=\textwidth]{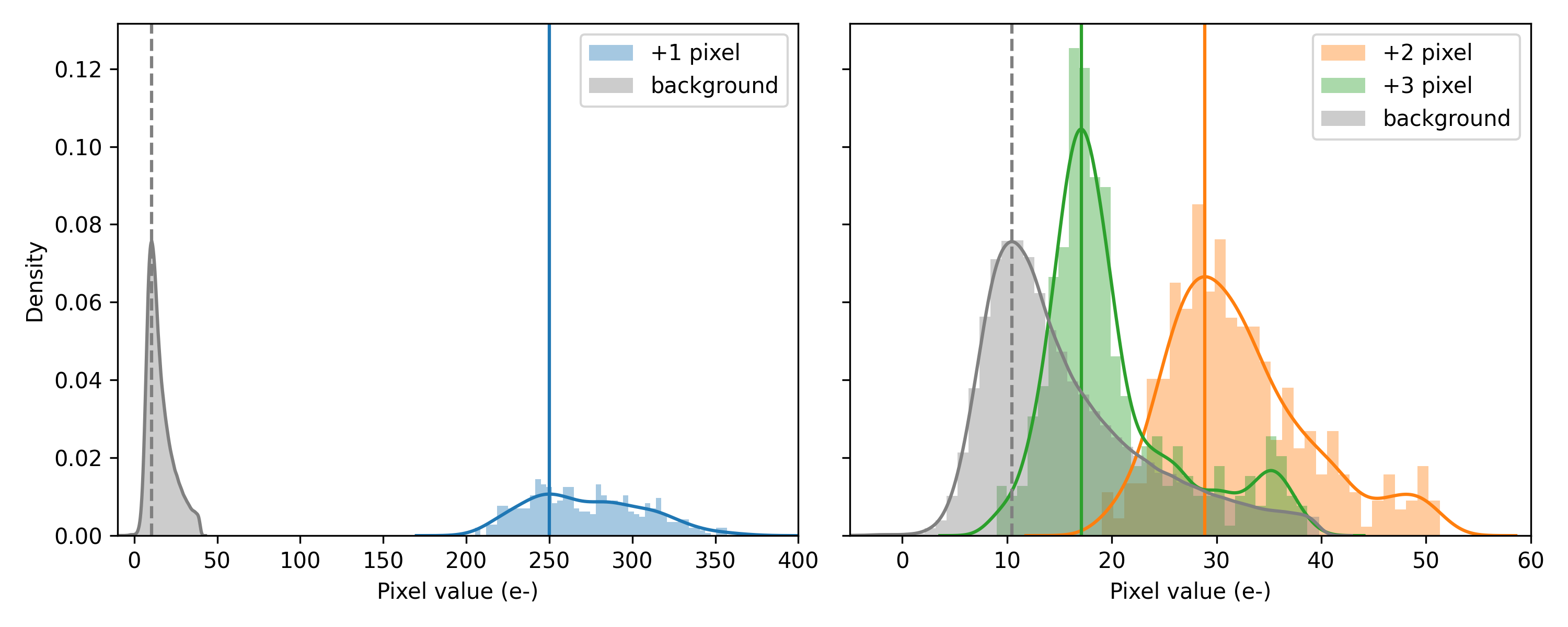}
    \caption{Example of kernel density estimation (KDE) peak-finding for the first three pixels in the serial CTE trail of 30k~e$^-$ hot pixels. The hot pixels are from quadrant B in the 2022 long dark stack. Histograms show the sigma-clipped distribution of pixel values of first three pixels in blue (+1), green (+2), and orange (+3). The grey histogram shows the background regions, and is the same in both panels. The curves and vertical lines show the KDE result and peak of the KDE for each distribution.}
    \label{kde}
\end{figure}

\begin{figure}[h!]
  \centering
  \includegraphics[width=\textwidth]{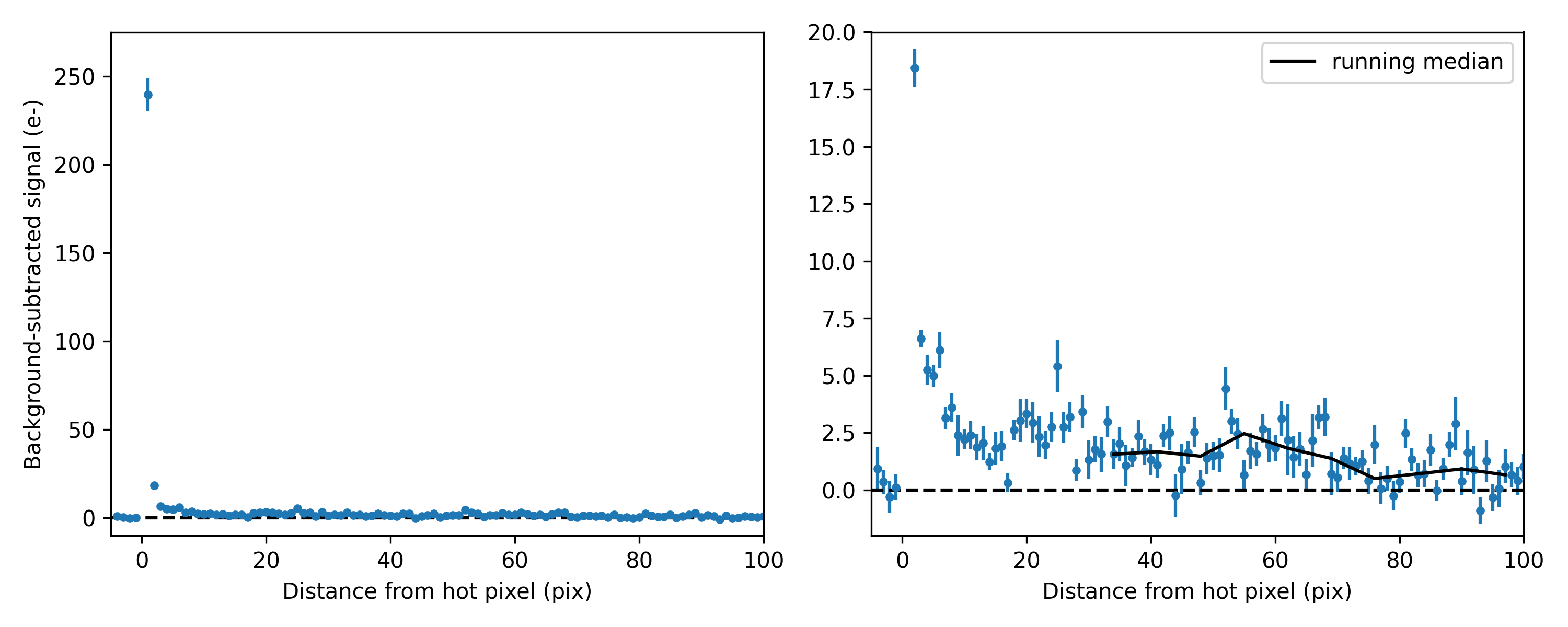}
    \caption{Serial CTE trail for 30k~$e^-$ hot pixels located in quadrant B of long dark stacks from 2022. The left panel shows the full trail, excluding the hot pixel. The right panel zooms in on the second pixel and beyond to better show the scatter in the trail. The dashed black line marks zero signal. A running median of the last 70 pixels (solid black line) shows that the serial trail still contains $\sim$0.5-1.0~e$^-$/pixel 100 pixels beyond the hot pixel.}
    \label{ampb_trail}
\end{figure}

\begin{figure}[h!]
  \centering
  \includegraphics[width=0.7\textwidth]{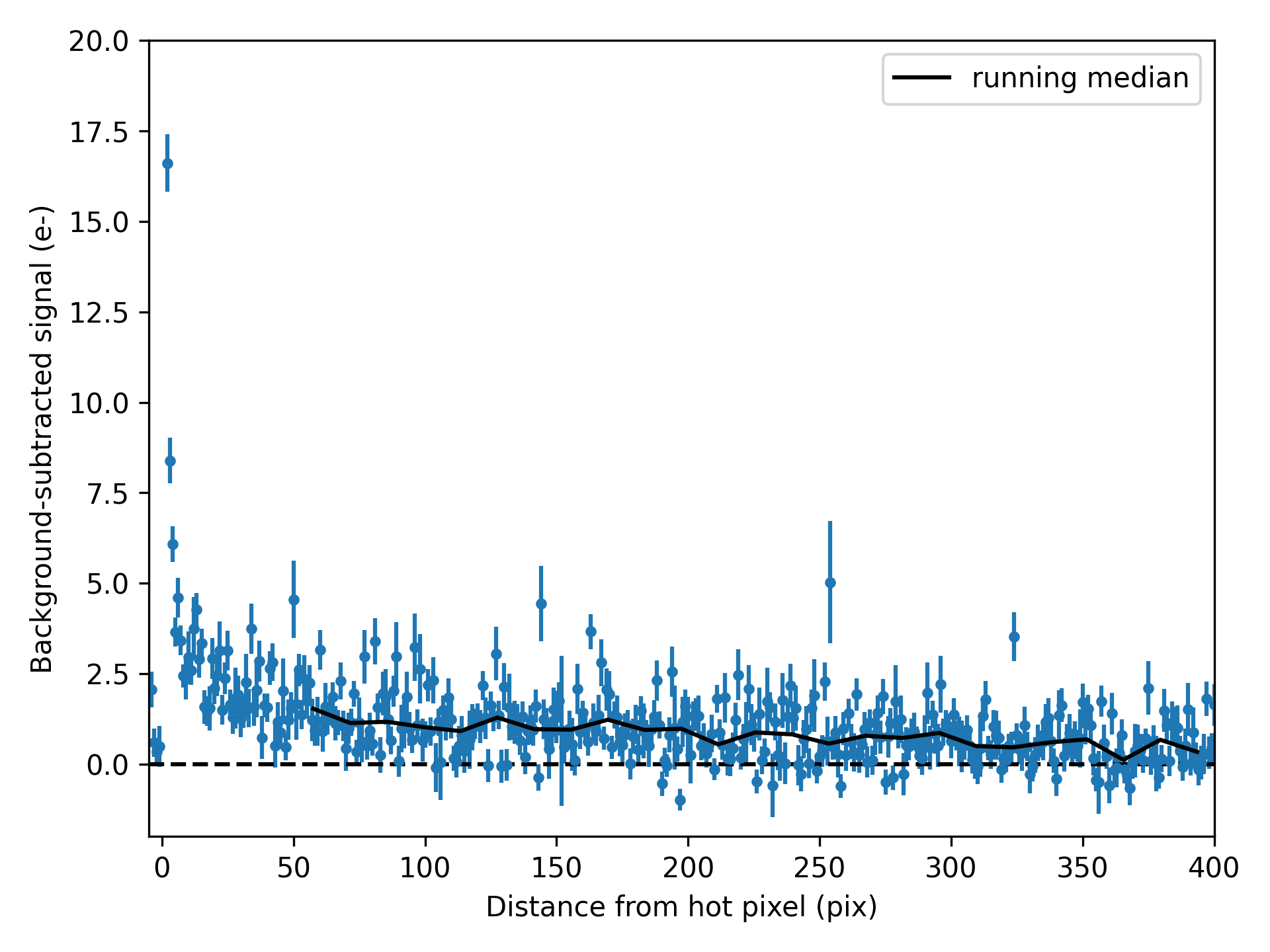}
    \caption{Same as the right-hand panel of Figure~\ref{ampb_trail} for 400 trailing pixels following 25k-75k~$e^-$ hot pixels that experience 1000-1600 serial transfers.}
    \label{extended_trail}
\end{figure}

An example of KDE peak-finding for the first three pixels in the serial trail of 30k~e$^-$ hot pixels in quadrant B from 2022 is given in Figure~\ref{kde}. The left panel shows the sigma-clipped distribution for the first pixel in the serial trail in blue. The curve and vertical line are the KDE fit and peak, respectively. The right panel shows the same for the second and third pixels in orange and green, respectively. Both panels show the same for the background regions in gray. The first pixel in the trail contains $\sim$1\% of the signal in the hot pixel, and is significantly brighter than subsequent pixels (0.1\% and 0.06\% for the second and third pixels, respectively). A similar level of signal is found in the trials for quadrants A and C, but quadrant D shows slightly lower levels: $\sim$0.6\% in the first, 0.06\% in the second, and 0.05\% in the third pixel. 

The serial CTE trail for 30k~e$^-$ hot pixels in quadrant B from 2022, with the background KDE peak value subtracted, is shown in Figure~\ref{ampb_trail}. Errors for each pixel value in the trail and the background are estimated by resampling the KDE fit with replacement, adding in random noise from the gaussian kernel bandwidth, and rerunning the KDE peak-finding routine 1,000 times. The error bars in Figure~\ref{ampb_trail} are the standard deviation of the resampled peaks of the trail pixel distributions combined in quadrature with the standard deviation of the resampled peaks of the background pixel distributions. The left panel shows the full trail, excluding the hot pixel. The right panel excludes the first trail pixel to show the rest of the trail in greater detail. The right panel also shows a running median of the last 70 pixels. Despite significant scatter, the serial trail remains above the background level by $\sim$0.5-1.0~e$^-$ at 100 pixels from the hot pixel.

The serial CTE trail may actually extend for several hundred pixels or more, like that of WFC3/UVIS \citep{anderson2024}. To detect and follow the trail for many more pixels than in Figure~\ref{ampb_trail}, we must select a wider range of hot pixels from closer to the center of the detector. In Figure~\ref{extended_trail}, we show the 400 trailing pixels for 25k-75k~e$^-$ hot pixels that experience 1000-1600 serial transfers. While the running median remains above zero ($\sim$0.3e$^-$/pixel) out to 400 pixels, there is significant scatter. It is also possible that residual dark current background or inadequately corrected bias shift \citep{golimowski2012} may be artificially inflating this extended trail. Considering that the first 100 pixels in this trail contain $\sim$470~e$^-$ and the following 300 pixels contain $\sim$250~e$^-$, we do capture the majority of the deferred charge within 100 pixels. Substantially increasing the model's trail length may also increase processing time for the serial CTE correction. For these reasons, we limit the serial trail length considered in this work to 100 pixels. Re-determining the model parameters described below with a longer serial trail is left to future work.

\subsection{Initial Trail and Trap Density Profiles} \label{initial_profiles}

\begin{figure}[t!]
  \centering
  \includegraphics[width=\textwidth]{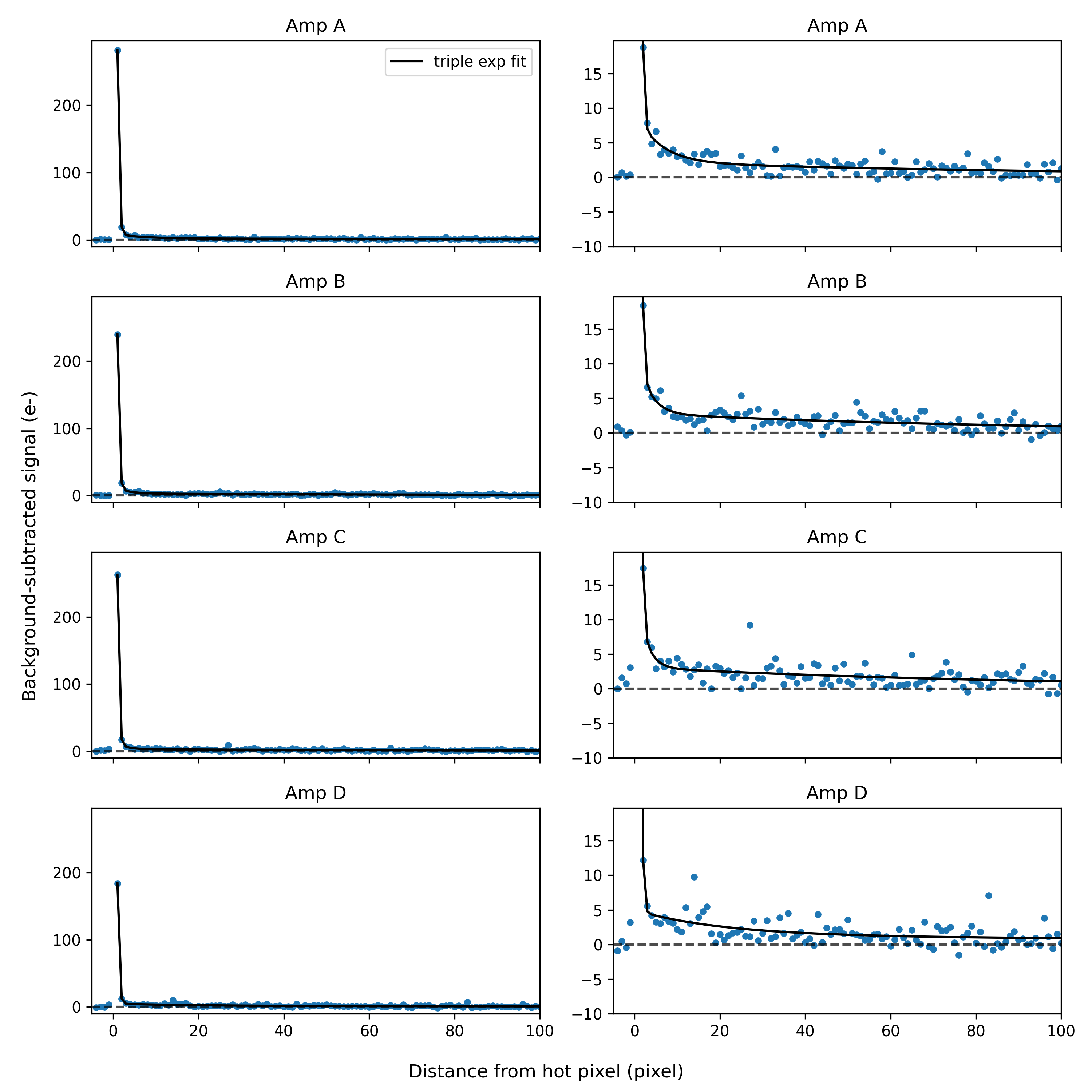}
    \caption{Same as Figure~\ref{ampb_trail} for all four quadrants of long dark stacks from 2022. The triple exponential fits (black line) describe the trail shapes well. Hot pixels in quadrant D appear to have fainter trails, particularly the first two pixels in the trail, than the other three quadrants.}
    \label{fit_trail}
\end{figure}

We use the 100-pixel trail of 30k~e$^-$ hot pixels to make an initial estimate of the serial trap density and trail profiles in each serial register. First, the trail from each quadrant is fit with a triple exponential, allowing amplitudes and decay factors of each exponential to vary. This represents the expected charge release behavior for three species of charge traps, which were found to fit the parallel CTE trails well by \cite{massey2010b} and \cite{massey2014}. Figure~\ref{fit_trail} shows the 30k~e$^-$ trails and triple exponential fits for each quadrant for long dark stacks from 2022. The shapes and overall amplitudes are similar, though quadrant D's trail appears to be fainter than the others, particularly in the first two pixels after the hot pixel.

The fit is summed to find the number of traps encountered by hot pixels during serial readout. Because the hot pixels are located at $x=1700\pm200$, we scale the sum of the fit and the fit itself by $2048/1700$ to estimate the total number of traps encountered by a charge packet traveling through all 2048 pixels of the serial register, and the brighter trail that would result. 

\begin{figure}[t]
  \centering
  \includegraphics[width=0.7\textwidth]{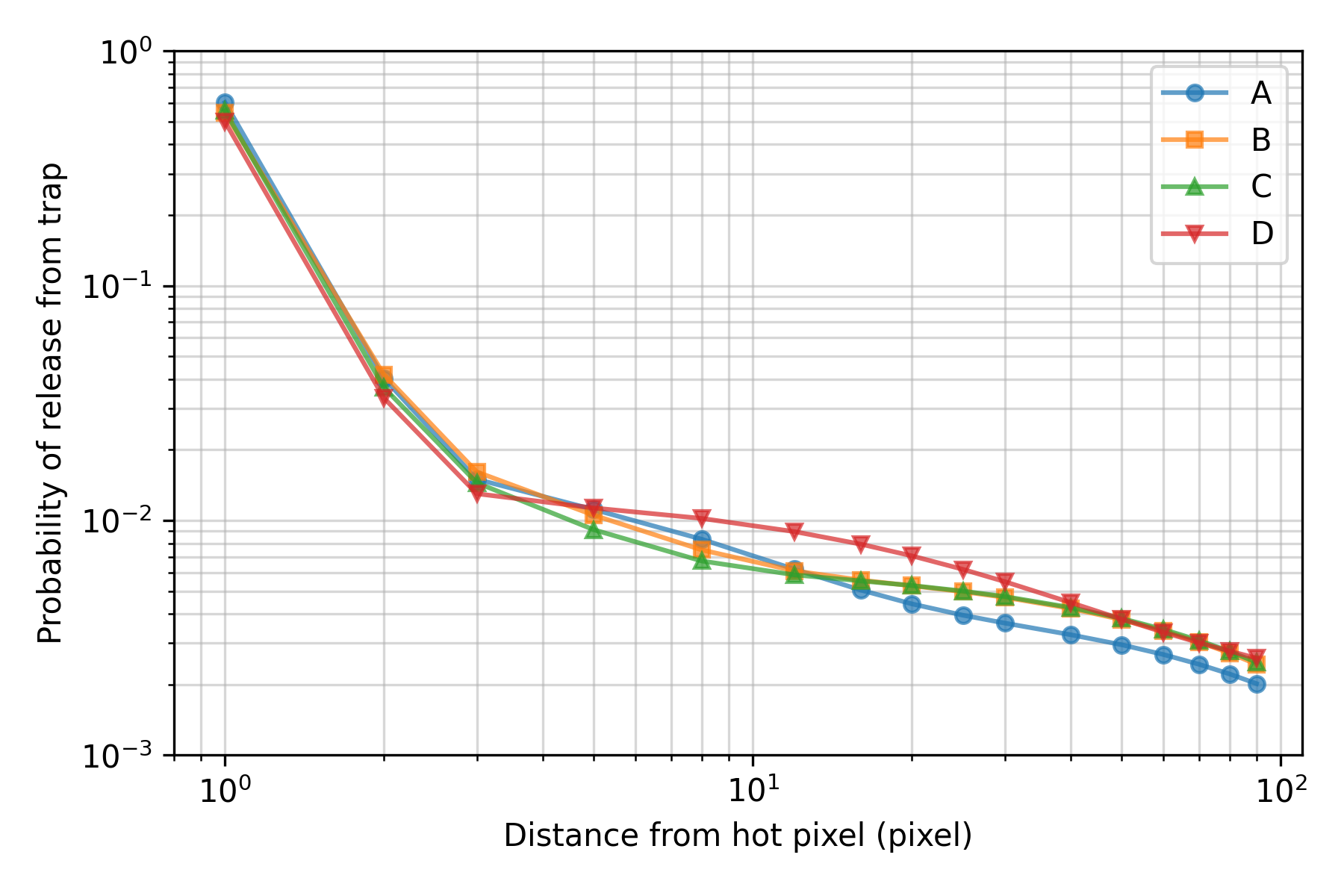}
    \caption{The initial trail profiles for the four quadrants, as determined from long dark stacks from 2022. The trail profile is the probability of release from charge traps as a function of distance from the hot pixel. The points show the set of pixel locations used to manually adjust the model, discussed in Section~\ref{adjust}.}
    \label{initial_trail_prof}
\end{figure}

The initial trail profile is simply the scaled pixel values in the fitted trail divided by the sum of these values, which represents the probability of charge release as a function of distance from the hot pixel. The initial trail profiles for all four quadrants are shown in Figure~\ref{initial_trail_prof}. The points are the pixel locations at which the parameter set is ``pinned'' in order to easily adjust the trail shape later in our analysis. The likelihood of trapped charged being released in one pixel transfer is 50-60\%, and 0.2-0.3\% after 90 pixel transfers, depending on the amplifier. Serial CTE trails are therefore much sharper than parallel trails, which have a $\sim$20\% likelihood of charge release in the first pixel.

\begin{figure}[t]
  \centering
  \includegraphics[width=0.7\textwidth]{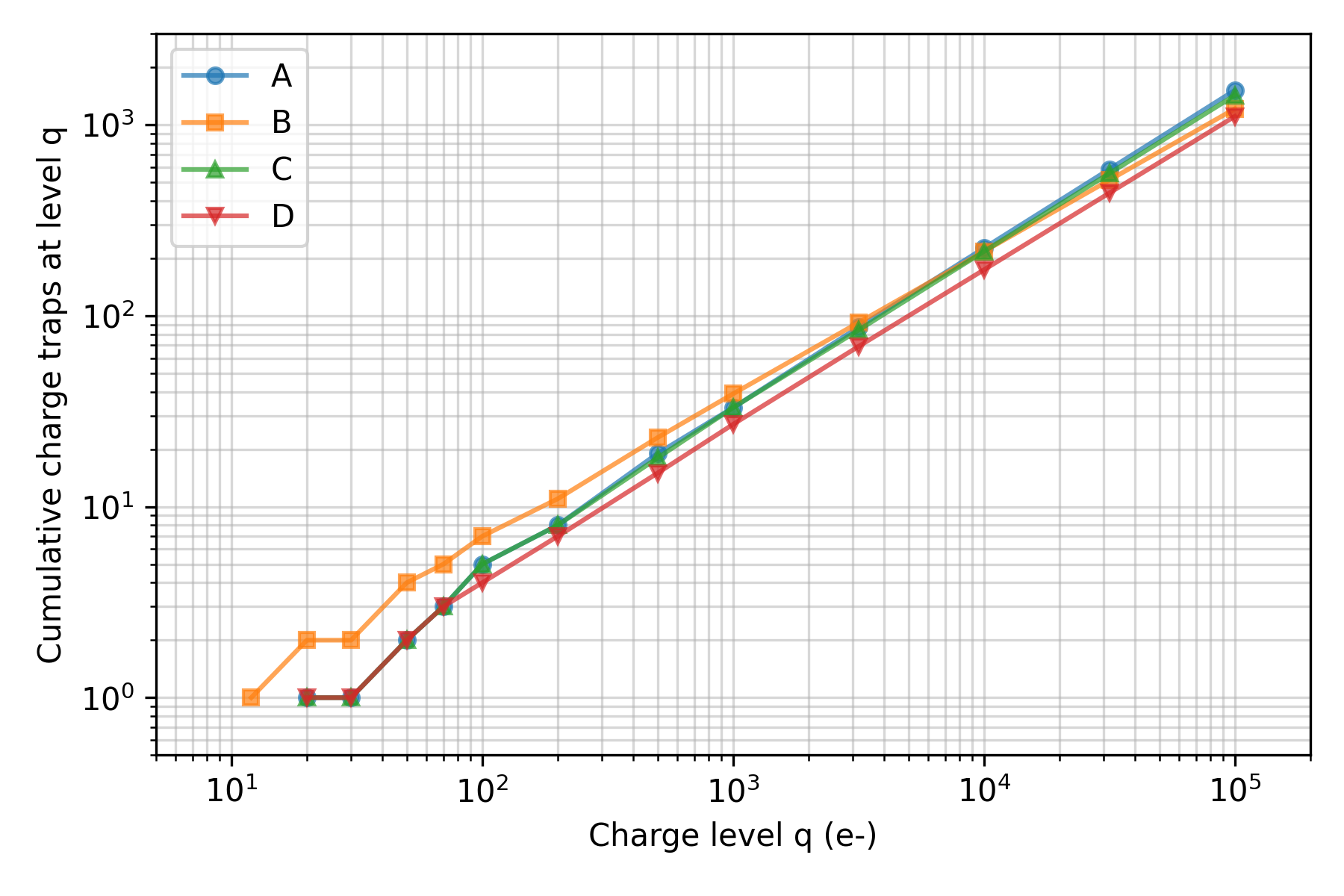}
    \caption{The initial trap density profiles for the four quadrants determined from long dark stacks from 2022. The trap density profile shows the total number of charge traps that affect a charge packet of size $q$ transferring from the edge of the quadrant furthest from the amplifier - 2048 transfers. The points are the set of q values, $\sim$logarithmically spaced, at which the profile is manually adjusted in Section~\ref{adjust}.}
    \label{initial_trap_density}
\end{figure}

To estimate the initial trap density profile for each serial register, we extract the trails at five pinning points in charge packet size: $q = 500$, 1000, 3160, 10k, 31.6k~e$^-$. The hot pixels used for each trail are within 20\% of the $q$ values listed and are located at $x=1700\pm200$. The first pixel in each trail is scaled for 2048 pixel transfers and divided by the probability of charge release in the first pixel (the first pinning point in Figure~\ref{initial_trail_prof}). This is an estimate of the total number of traps encountered by a charge packet of size $q$ transferring through the entire serial register. The total traps are added to the $q$ levels as a first-order accounting for charge lost in serial trails. A power-law fit to the total traps as a function of $q$ is performed and then extrapolated to smaller charge packets. The result is the initial trap density profile, which is plotted for all four quadrants in Figure~\ref{initial_trap_density}. 

The points are the charge packet sizes at which the parameter set is ``pinned'' for ease of manual adjustment: $q = 12$, 20, 30, 50, 70, 100, 200, 500, 1000, 3160, 10k, 31.6k, 100k~e$^-$. For serial CTE in ACS/WFC, charge packets must be fairly large before traps begin to affect them. This may be because the trap capture time constant is longer than the serial dwell time for small $q$, as discussed in Section~\ref{cte_background}. This is in contrast to parallel CTE, where there are more traps than electrons for $q\leq200$~e$^-$ \citep{anderson2018}. The initial trap density profiles are similar among quadrants. Quadrants B and D have slightly shallower profiles than quadrants A and C, and quadrant B appears to have more traps at the low $q$ end.

\subsection{Adjustments \& Final Profiles} \label{adjust}

\begin{figure}[th!]
  \centering
  \includegraphics[width=0.8\textwidth]{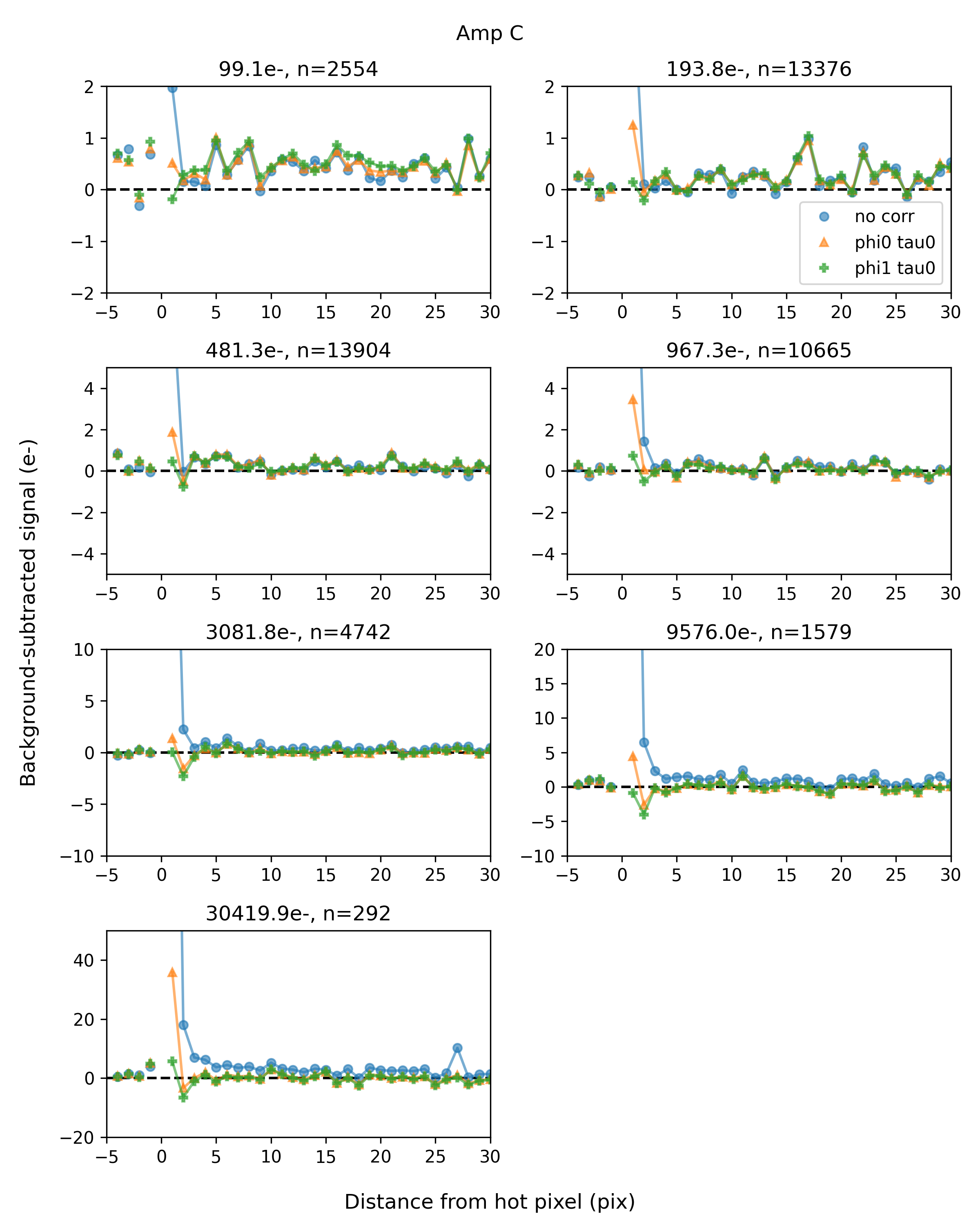}
    \caption{First 30 pixels of the serial CTE trails in quadrant C of uncorrected darks (blue points), darks corrected with the initial profiles (orange triangles), and darks corrected with the first iteration of manually-adjusted trap density profiles (green crosses). The dashed black line marks zero signal. In the legend notation, ``phi'' refers to the trap density profile and ``tau'' to the trail profile, so ``phi1 tau0'' means the trap density profile has been manually-adjusted once, and the trail profile has not. Each panel shows a bin in hot pixel brightness: $q = 100$, 200, 500, 1000, 3160, 10k, 31.6k~e$^-$. The median hot pixel value and the number of pixels in each bin are shown above each panel. The first pixel in the uncorrected trails are usually too bright to view the rest of the trail in detail, and is therefore typically excluded from each panel.} 
    \label{first_iters_trails}
\end{figure}

Next, the serial CTE correction is run on the 2022 long dark stacks using the initial trap density and trail profiles. We extract the serial CTE trails at seven charge packet sizes ($q = 100$, 200, 500, 1000, 3160, 10k, 31.6k~e$^-$) in the uncorrected and corrected long dark stacks. Plotted in Figure~\ref{first_iters_trails} are the first 30 pixels of the uncorrected trails (blue points) and corrected trails using the initial parameter sets (orange triangles) in an example quadrant (C). (The green crosses represent a further adjustment to the profiles, and are discussed below.) We show the median hot pixel value and the number of pixels in the bin at the top of each panel. The first pixel in the uncorrected trails is typically excluded from each panel in order to zoom in on the corrected trails.

If the serial CTE correction were perfect, the corrected trails in each panel would have zero signal. Clearly, serial CTE correction with the initial parameter set (orange trail) is a substantial improvement over no correction at all (blue trail). For many bins, the first pixel is still bright relative to the remainder of the trail, but it is significantly lower than in the uncorrected trail. 

For the bins that have positive-valued first pixels in the orange trails, we manually increase the total number of traps at those $q$ levels in the trap density profiles. We manually decrease the total traps if the first trail pixel is negative. The green trails in Figure~\ref{first_iters_trails} show the result of the first adjustment of the trap density profiles. In most panels, the first pixel is closer to zero, the second pixel is negative, and the third and further pixels are largely unaffected. The similarity in shape of the green trails across $q$ levels suggests that the trail profiles should be adjusted next. In particular, the probability of release into the second pixel should be decreased.

\begin{figure}[h!]
  \centering
  \includegraphics[width=0.8\textwidth]{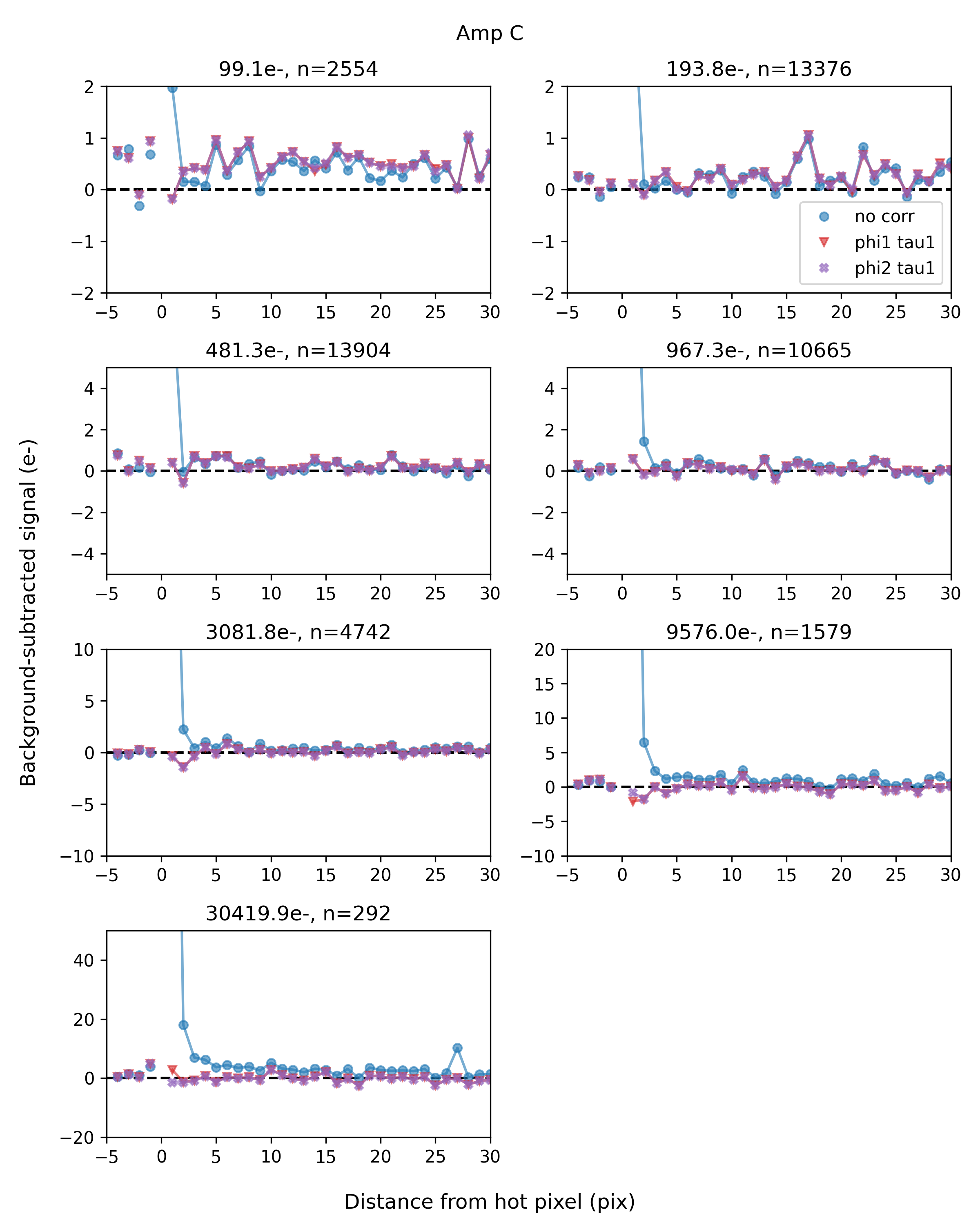}
    \caption{Same as Figure~\ref{first_iters_trails}, but for two further iterations of manual adjustment to the trail profiles (red triangles) and trap density profiles (purple Xs). The uncorrected trails are again shown with blue circles.}
    \label{final_iters_trails}
\end{figure}

\begin{figure}[h!]
  \centering
  \includegraphics[width=0.8\textwidth]{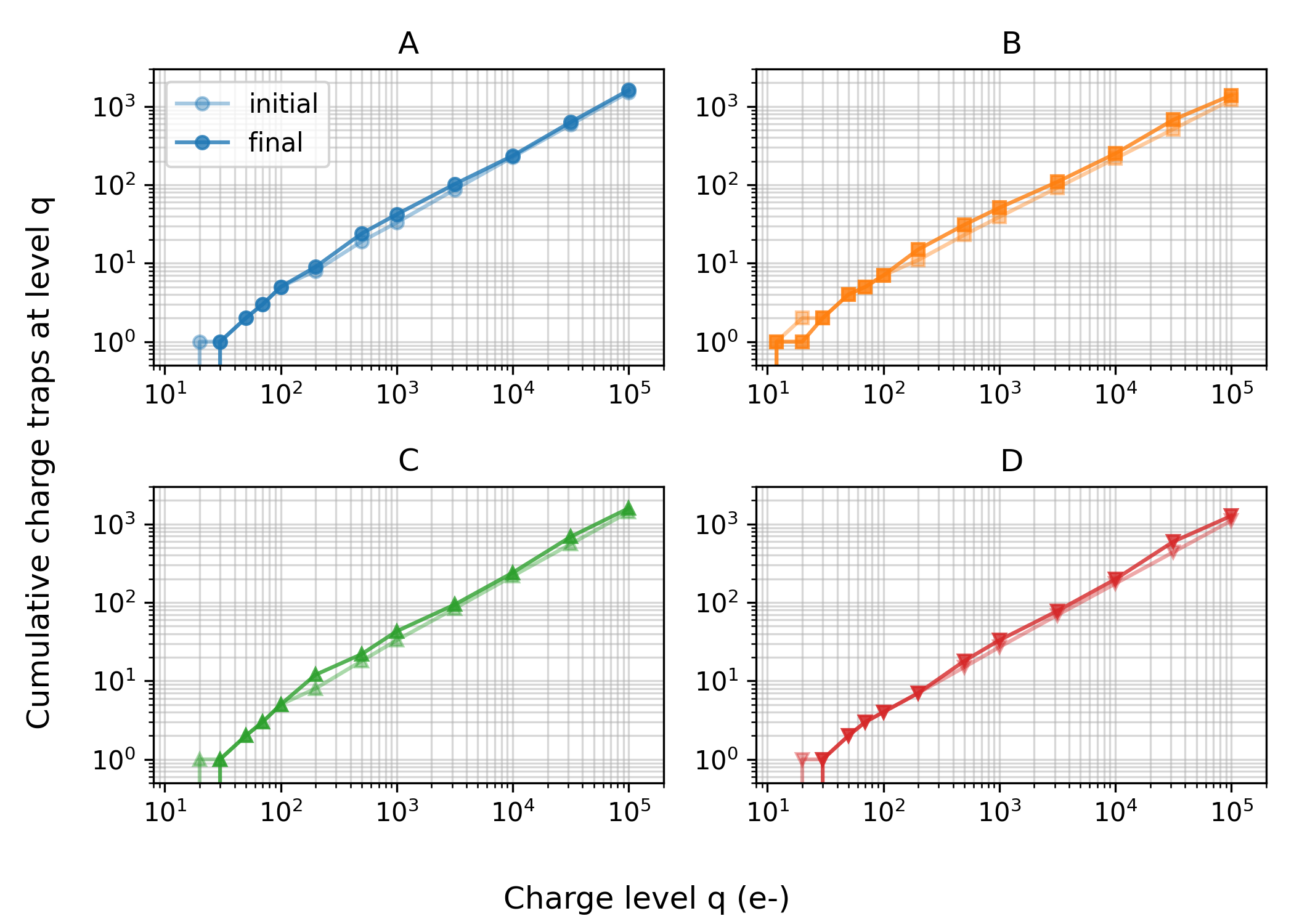}
    \caption{Same as Figure~\ref{initial_trap_density} for the final trap density profiles, separated by quadrant. The initial trap density profiles are also shown for comparison.}
    \label{final_trap_prof}
\end{figure}

\begin{figure}[h!]
  \centering
  \includegraphics[width=0.8\textwidth]{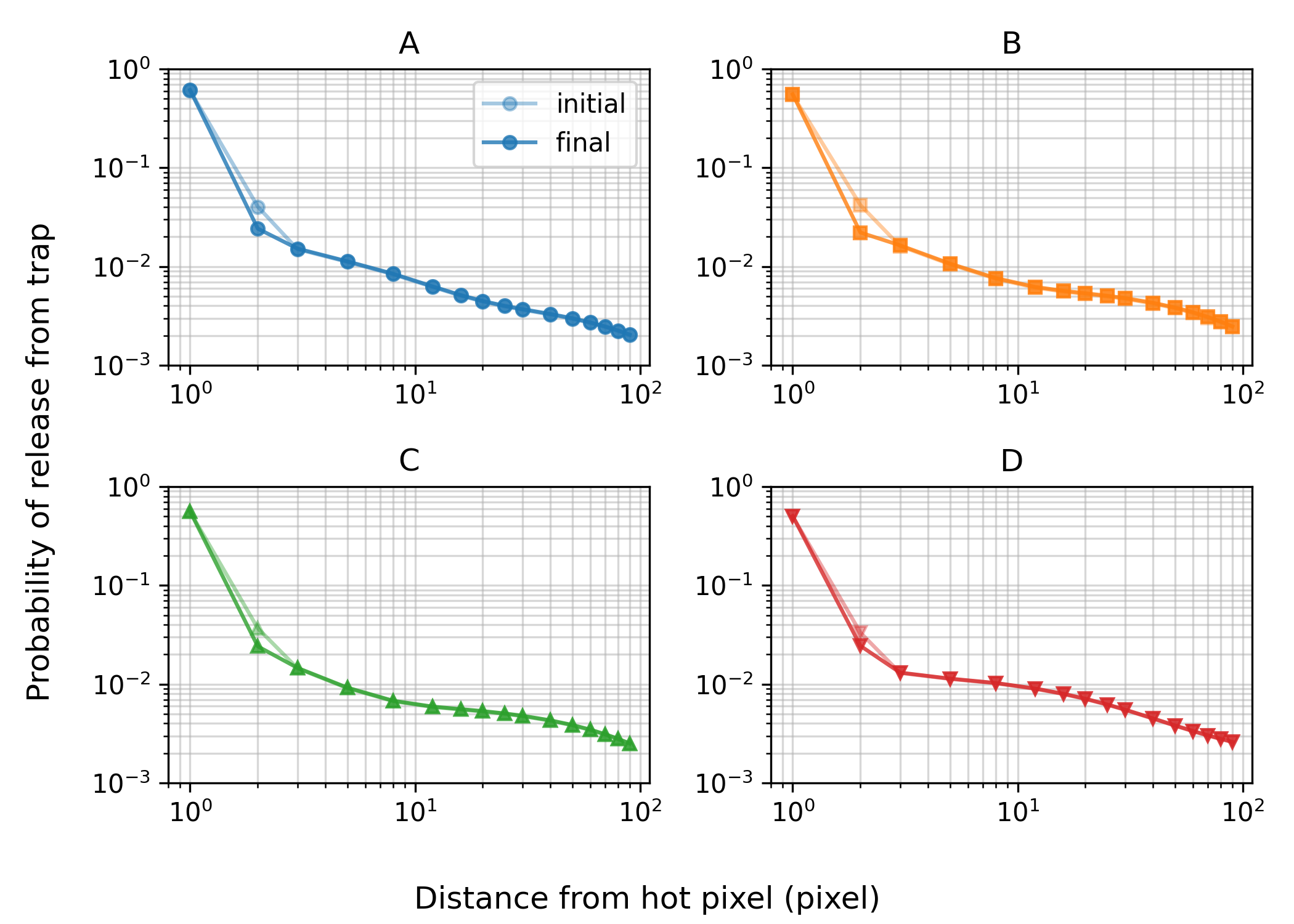}
    \caption{Same as Figure~\ref{initial_trail_prof} for the final trail profiles, separated by quadrant. The initial trail profiles are also shown for comparison.}
    \label{final_trail_prof}
\end{figure}

We adjust the trail profiles and show the resulting trails as red triangles in Figure~\ref{final_iters_trails}. The purple X's are the final serial CTE trails, which result from a final adjustment to the $q=10$k and 31.6k~e$^-$ levels in the trap density profiles for quadrant C, and a few other $q$ levels in the other quadrants. No further adjustments are made to the profiles when the excursions from zero in the first few pixels are similar to the variability in the four ``upstream'' pixels and pixels far from the hot pixel, i.e., when the corrected trails approximate background noise. While the final serial CTE trails are not perfectly flat in all bins, the improvement from the uncorrected trails is drastic. Similar improvements are seen in the quadrants not shown here.

In Figures~\ref{final_trap_prof} and \ref{final_trail_prof}, we show the final trap density and trail profiles, respectively, for each quadrant, and the initial profiles of each for comparison. Overall, the adjustments to the trap density profiles were relatively minor. The downward adjustments of the second pixel in the trail profiles are more substantial. Charge trapped from the first trail pixel may be released into the second pixel, i.e., ``re-trailing'', which could explain the initial overestimation of release chances in the second pixel. In Tables~\ref{table_trap} and \ref{table_trail}, we provide the final trap density profile data and trail profile data, respectively.

\begin{table}[t]
\small
\centering
\caption{\textbf{Trap density profiles for each WFC quadrant:} Cumulative charge traps encountered by a given size charge packet $q$ during 2048 pixel transfers.}
\begin{tabular}{c|cccc}
\toprule
\textbf{Charge level} & \multicolumn{4}{c}{\textbf{Cumulative charge traps}} \\
 $q$ (e~$^-$) & \textbf{A} & \textbf{B} & \textbf{C} & \textbf{D} \\
\hline
12 & 0 & 1 & 0 & 0 \\
20 & 0 & 1 & 0 & 0 \\
30 & 1 & 2 & 1 & 1 \\
50 & 2 & 4 & 2 & 2 \\
70 & 3 & 5 & 3 & 3 \\
100 & 5 & 7 & 5 & 4 \\
200 & 9 & 15 & 12 & 7 \\
500 & 24 & 31 & 22 & 18 \\
1000 & 42 & 51 & 43 & 33 \\
3160 & 102 & 109 & 94 & 78 \\
10000 & 235 & 250 & 240 & 198 \\
31600 & 631 & 676 & 689 & 597 \\
99999 & 1616 & 1389 & 1593 & 1273 \\
\bottomrule
\end{tabular} 
\label{table_trap}
\end{table}

\begin{table}[t]
\small
\centering
\caption{\textbf{Trail profiles for each WFC quadrant:} Probability of charge release as a function of distance from the hot pixel.}
\begin{tabular}{c|cccc}
\toprule
\textbf{Distance} & \multicolumn{4}{c}{\textbf{Probability of release}} \\
 (pixels) & \textbf{A} & \textbf{B} & \textbf{C} & \textbf{D} \\
\hline
1 & 0.6135 & 0.5555 & 0.5627 & 0.5040 \\
2 & 0.0242 & 0.0221 & 0.0242 & 0.0244 \\
3 & 0.0152 & 0.0164 & 0.0146 & 0.0131 \\
5 & 0.0113 & 0.0107 & 0.0092 & 0.0114 \\
8 & 0.0084 & 0.0076 & 0.0068 & 0.0103 \\
12 & 0.0063 & 0.0062 & 0.0059 & 0.0090 \\
16 & 0.0051 & 0.0057 & 0.0056 & 0.0080 \\
20 & 0.0045 & 0.0054 & 0.0053 & 0.0071 \\
25 & 0.0040 & 0.0051& 0.0051 & 0.0062 \\
30 & 0.0037 & 0.0048 & 0.0048 & 0.0055 \\
40 & 0.0033 & 0.0043 & 0.0043 & 0.0045 \\
50 & 0.0030 & 0.0038 & 0.0039 & 0.0038 \\
60 & 0.0027 & 0.0034 & 0.0035 & 0.0034 \\
70 & 0.0025 & 0.0031 & 0.0031 & 0.0030 \\
80 & 0.0023 & 0.0028 & 0.0028 & 0.0028 \\
90 & 0.0020 & 0.0025 & 0.0025 & 0.0026 \\
\bottomrule
\end{tabular} 
\label{table_trail}
\end{table}

\begin{figure}[t]
  \centering
  \includegraphics[width=\textwidth]{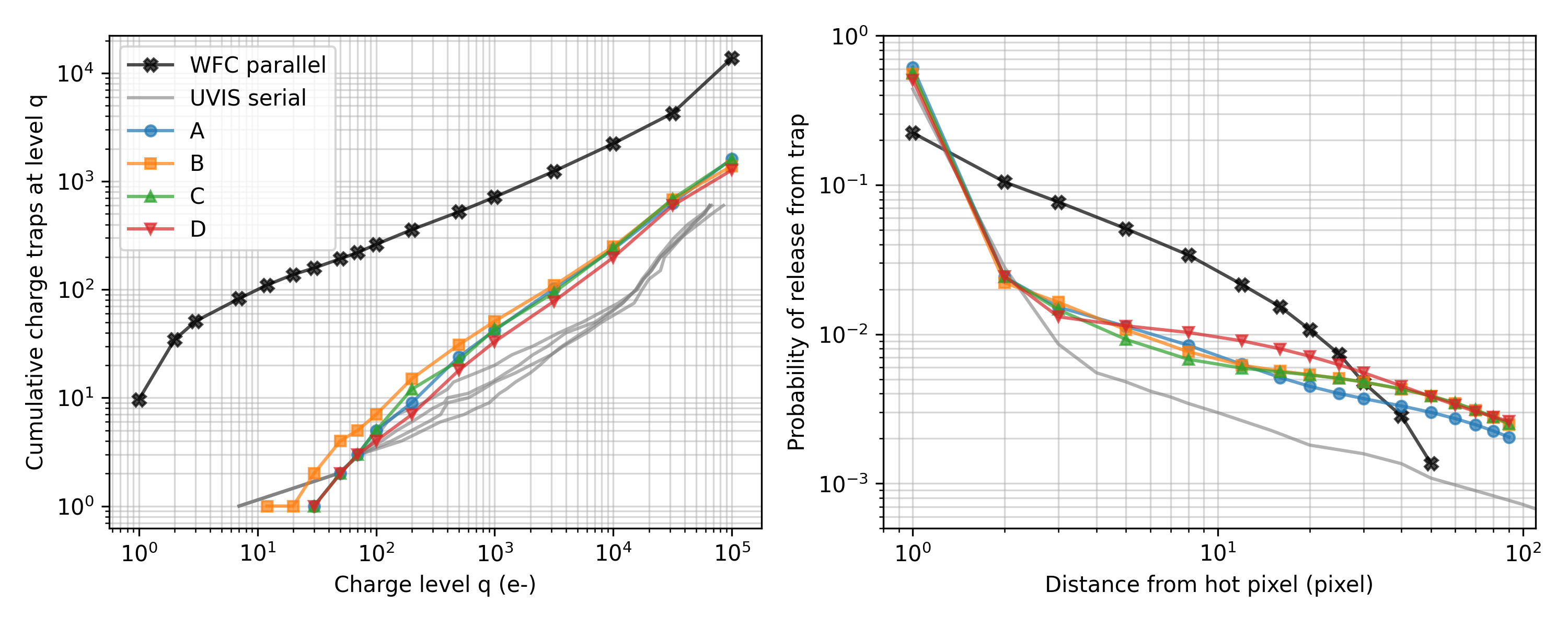}
    \caption{Trap density (left) and trail profiles (right) for ACS/WFC serial CTE (same as in Figures~\ref{final_trap_prof} and \ref{final_trail_prof}), ACS/WFC parallel CTE (black Xs), and WFC3/UVIS serial CTE (gray curves).}
    \label{profile_comparison}
\end{figure}

Figure~\ref{profile_comparison} shows the final serial CTE trap density and trail profiles determined in this work alongside those for parallel CTE in ACS/WFC \citep{anderson2018} and serial CTE in WFC3/UVIS \citep{anderson2024}. The trap density profile for WFC parallel CTE has been scaled to mid-2022, which corresponds to the epoch of the WFC serial CTE profile. The epoch of the UVIS trap density profile is mid-2023 \citep{anderson2024}, near enough in time to that of the WFC for comparison purposes. Clearly, many more charge traps affect charge packets during parallel transfers than during serial transfers in WFC. The parallel profile has a much shallower slope than the serial profile, and traps actually outnumber electrons in small charge packets ($q\lesssim500$~e$^-$). The parallel trail is much less sharp than the serial trail, such that the first pixel contains $\sim$20\% of the deferred charge, but subsequent pixels are much brighter than those in the serial trail. The parallel trail also becomes undetectable by $\sim$60~pixels, whereas the serial trail extends further.

UVIS has fewer traps for charge packets $q\gtrsim10$~e$^-$ than WFC in the serial direction, and an interesting upturn in number of traps above $q>10k$~e$^-$ that is not reflected in WFC. The UVIS serial trail is similar in shape to that of WFC, but it is fainter beyond two pixels from the hot pixel, and extends for 2103 pixels, as discussed in \citep{anderson2024}. This reinforces the notion that the WFC serial trail  could extend further than included in our current model, which may be explored in future work.

\subsection{Time-Dependence of Serial CTE} \label{time_dependence}

\begin{figure}[t]
  \centering
  \includegraphics[width=0.7\textwidth]{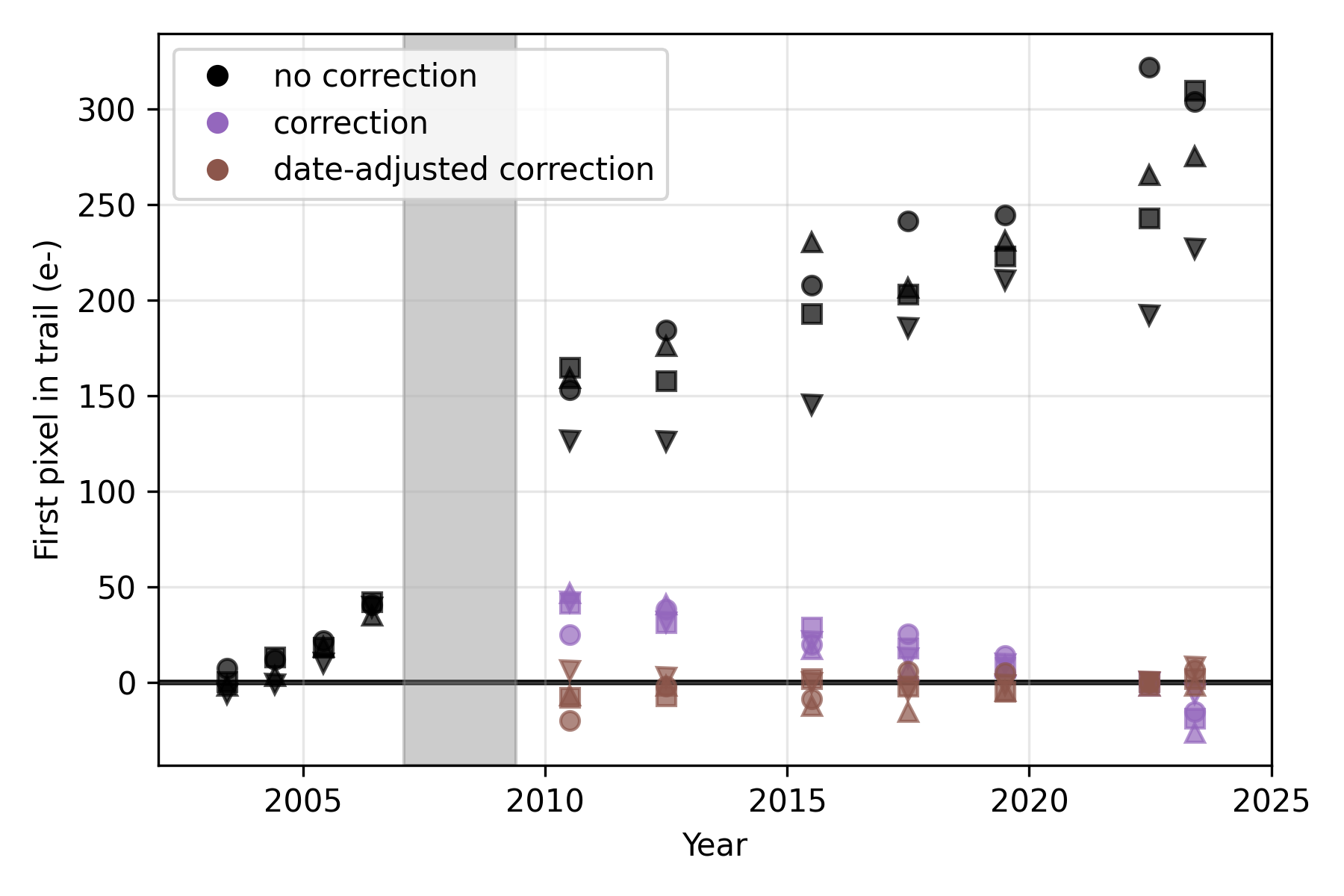}
    \caption{First trail pixel values as a function of time for $31.6$k~e$^-$ hot pixels. The symbols represent the four quadrants as in Figures~\ref{initial_trail_prof} and \ref{initial_trap_density}. The black points are for uncorrected darks, the purple points are for darks corrected with the final profiles shown in Figures~\ref{final_trap_prof} and \ref{final_trail_prof}, and the brown points are for darks corrected with the final profiles and manually-adjusted time constants. The period when ACS/WFC was offline is shown by the gray region. The black line marks zero signal.}
    \label{first_pix}
\end{figure}

The final trap density and trail profiles shown in Figures~\ref{final_trap_prof} and \ref{final_trail_prof} were determined using 2022 data. We run the serial CTE correction on datasets from several years to test the time-dependence in the model. Figure~\ref{first_pix} shows the signal in the first pixel in the trails from each quadrant for the largest charge packet size studied, $q=31.6$k~e$^-$, as a function of time. We use the residual signal in the first trail pixel as a proxy for residual signal in the entire trail, since the first trail pixel contains most of the deferred charge. The black points are from uncorrected darks and the purple points are from darks corrected with the final profiles discussed in Section~\ref{adjust}. The symbols correspond to the detector quadrants as in Figures~\ref{initial_trail_prof} and \ref{initial_trap_density}.

Figure~\ref{first_pix} shows that prior to SM4, serial CTE losses were less than a third of post-SM4 losses and therefore much less of a concern. We restrict further analysis to post-SM4 data and also choose to skip the serial CTE correction step for pre-SM4 data in CALACS (further discussed in Section~\ref{implementation}). Additionally, there is non-linear behavior in the evolution of first pixel residuals when comparing the pre- and post-SM4 time periods. This may indicate that the shape of the serial CTE trails changed, perhaps as a result of the temperature change in 2006 or with the replacement of the WFC electronics during SM4. 

For post-SM4 but prior to 2022, there is clearly residual signal in the first trail pixel in the corrected data (purple points). In fact, the residual signal seems to decrease linearly as a function of time toward 2022. In 2023 data, serial CTE losses appear to be overcorrected. This suggests that adjustments of the time constants $t_0$ and $t_1$ are needed, as discussed in Section~\ref{model}. Allowing $t_0$ to remain as the MJD of the installation date of ACS on HST, as it is for the parallel CTE correction, and setting $t_1$ to the average MJD of the 2022 data does not reduce the residual signal to zero. Through trial and error, we find that $t_0 = \mathrm{MJD}\ 49000$ (Jan 1993) and $t_1 = \mathrm{MJD}\ 60500$ (July 2024) result in the best time-dependent correction for all post-SM4 darks studied, as shown by the brown points in Figure~\ref{first_pix}. With these time constants, the post-SM4 trap density scale factor is lower overall and grows more slowly than with the original time constants discussed in Section~\ref{model}. There is some scatter in the first pixel residuals among the quadrants from data prior to 2022, but taken together, the residuals are closest to zero given these time constants. Similar trends were seen for dimmer hot pixels, so the adjusted time constants are effective at reducing the trail residuals for hot pixels of any brightness.

\section{Results} \label{results}

\subsection{Qualitative Assessment of Correction} \label{qual}

\begin{figure}[t]
  \centering
  \includegraphics[width=\textwidth]{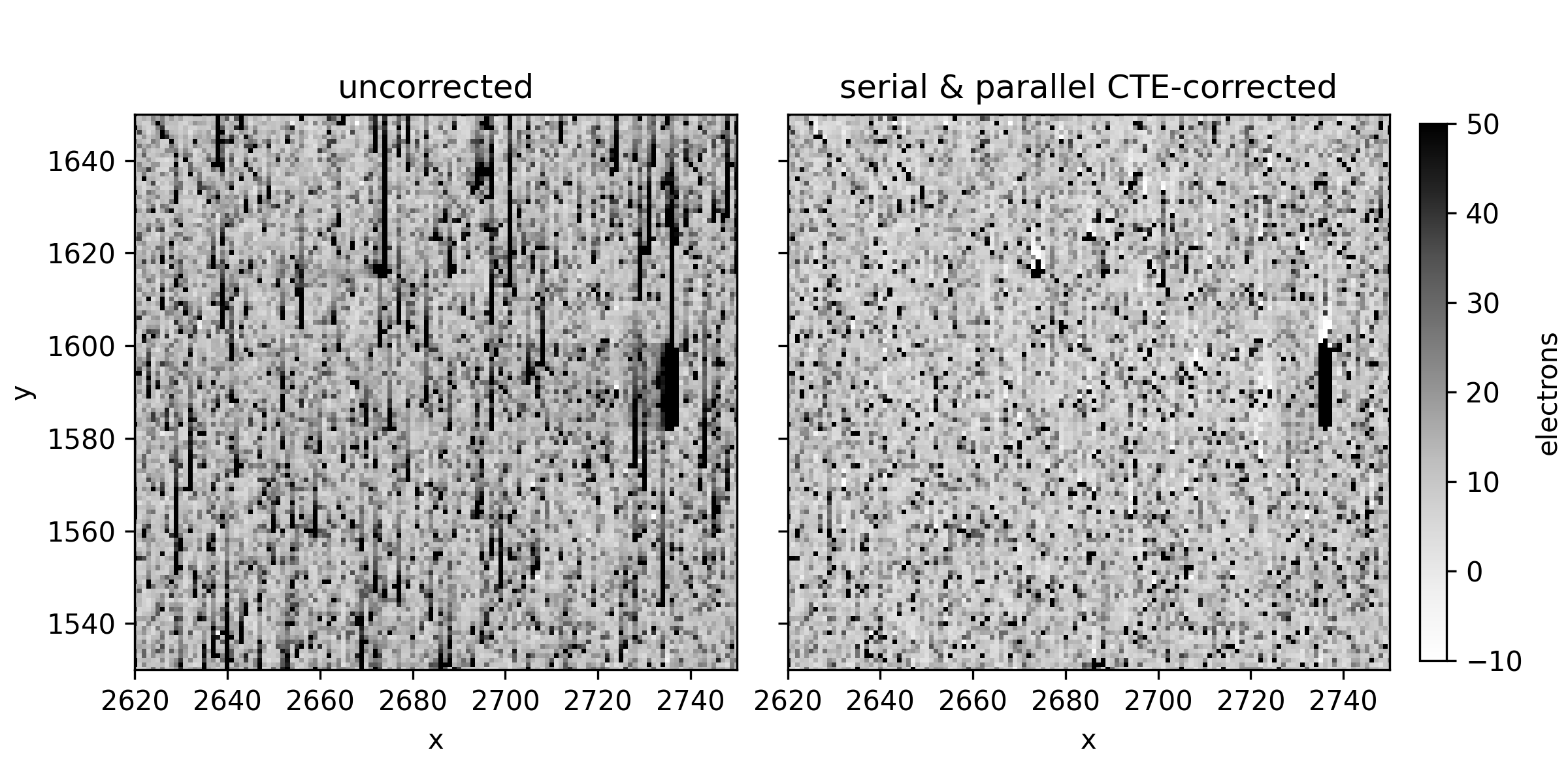}
    \caption{Same as Figure~\ref{trail_image} with an additional panel showing the serial- and parallel-CTE-corrected image.}
    \label{dark_corr_uncorr}
\end{figure}

In Figure~\ref{dark_corr_uncorr}, we show the long dark from Figure~\ref{trail_image} in the left panel and the serial- and parallel-CTE-corrected version in the right panel. The CTE corrections provide a substantial improvement to the data quality, which is visually evident. The serial and parallel trails are almost entirely removed, and most warm and hot pixels now appear as individual pixels, rather than smears of several pixels in both directions.

\subsection{Impact on Observations of Stars} \label{impact}

It is difficult to visually discern the effects of serial CTE on images of stellar fields, but we can quantify its impact. Data from two calibration programs are studied. Calibration program 14881 from February 2017 (PI Anderson) obtained four pairs of long (700s) and short (35s) exposures of a stellar field in the Milky Way bulge. Exposures were taken in the F606W filter and WFC1B-2K subarray. The short exposures were post-flashed to a range of background levels: 21, 49, 77, and 104~e$^-$. The purpose of this program is to compare photometry of stars present in both the short and long exposures to evaluate how well the CTE correction performs as a function of stellar brightness and background level. 

We also utilize a single 150s F606W exposure of 47 Tuc from calibration program 17336 from July 2024 (PI Chiaberge) for direct comparison of corrected and uncorrected data. The background level in this image is $\sim$15~e$^-$.

All images were fully processed from RAW with CALACS version 10.3.5, which includes bias, parallel CTE, dark, and flat correction, but does not include serial CTE correction. We refer to these data as ``uncorrected'' in what follows, due to the lack of serial CTE correction. Separately, all images were again fully processed from RAW with CALACS version 10.3.5, but the pixel-based serial CTE correction was included at the appropriate stage in processing (prior to parallel CTE correction). These are referred to as the ``corrected'' images. The corrected long exposures from CAL-14881 were combined with ACSREJ to increase signal-to-noise and reject cosmic rays\footnote{One long exposure was excluded from the combination due to out-of-family jitter excursions.}. 

With \texttt{hst1pass}, we obtained PSF-fitting photometry and precise positions of stars in the exposures \citep{anderson2022}. We used the spatially-variable PSF model for post-SM4 F606W data\footnote{\url{https://www.stsci.edu/~jayander/HST1PASS/LIB/PSFs/STDPSFs/ACSWFC/STDPSF_ACSWFC_F606W_SM4.fits}}. Perturbations to the PSF were not allowed during fitting, in order to avoid potentially compensating for serial CTE trails with the shape of the PSF. 

Because deeper images accumulate relatively more electrons for each star and more background signal than shorter images, they suffer from relatively less CTE loss in both transfer directions. The parallel- and serial-CTE corrections also reconstruct the lost signal more accurately for those reasons. We therefore treat the stellar photometry from the corrected long exposures as ``truth'' and scale the measured fluxes by the ratio of exposure times for comparison to the short exposure photometry. Saturated, poorly-fit, and potentially spurious sources were removed from the ``truth'' photometry catalog by imposing cuts on quality of fit, \texttt{q}, and sharpness, \texttt{C}:
\begin{itemize}
\item $0 < \mathrm{\texttt{q}} \leq 0.15$
\item $-0.03 < \mathrm{\texttt{C}} < 0.03$.
\end{itemize}
Then, sources in the ``truth'' and short catalogs were matched in $x$ and $y$ positions using the nearest-neighbor algorithm in \texttt{scipy.spatial.KDTree}. 

\begin{figure}[th]
  \centering
  \includegraphics[width=0.9\textwidth]{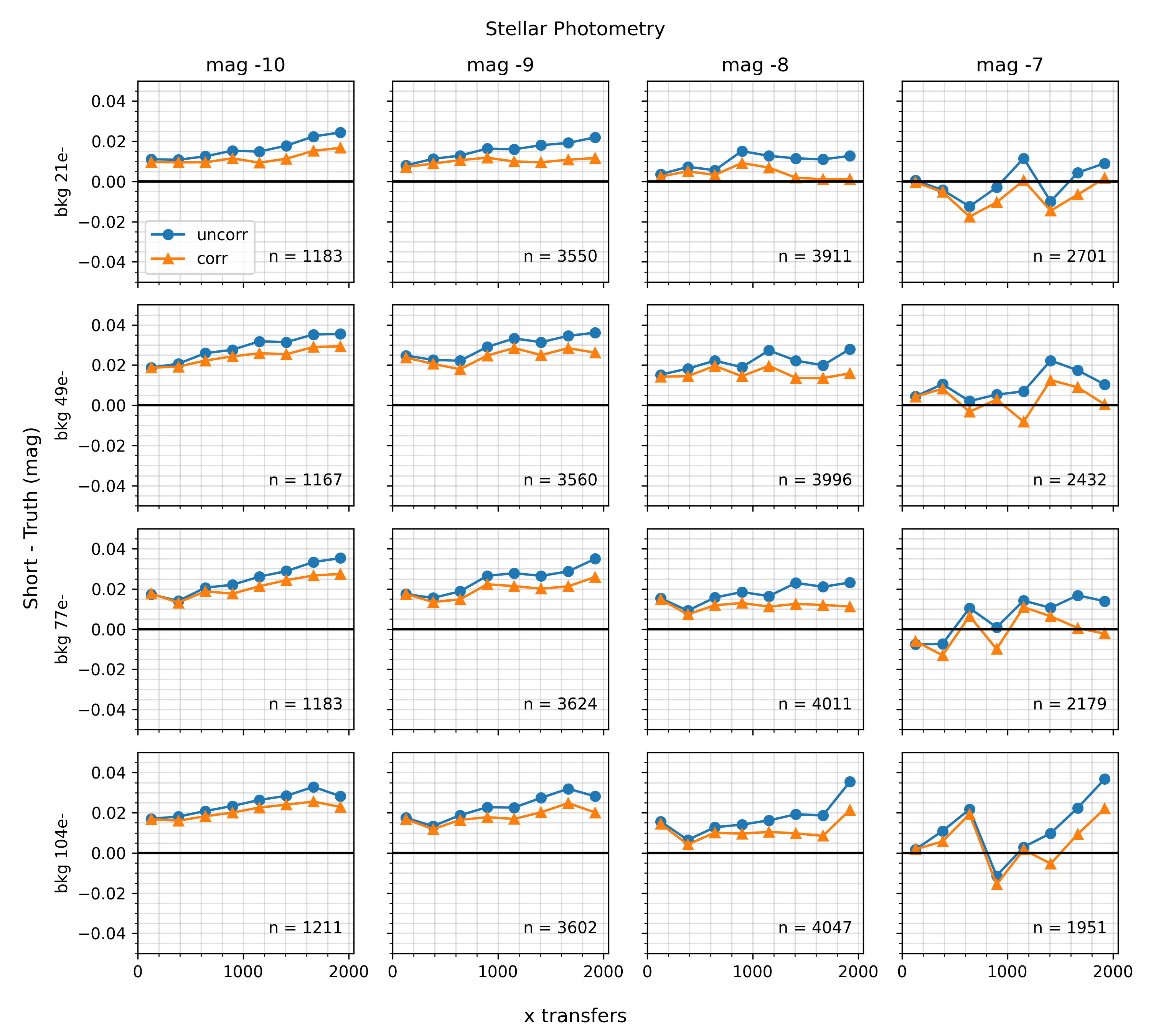}
    \caption{Magnitude differences between stars in short and ``truth'' exposures as a function of $x$ transfers. The panels represent bins in instrumental stellar magnitude (columns) and background level (rows) in the short exposures. Uncorrected short exposures are plotted as blue circles and corrected short exposures are plotted as orange triangles. Zero magnitude difference is shown by a black line for reference, and the number of stars in each panel is shown in the lower right corner.}
    \label{serial_phot_recovery}
\end{figure}

In Figure~\ref{serial_phot_recovery}, we plot magnitude differences between stars in the short and ``truth'' exposures as a function of $x$-direction pixel transfers. (This figure is the serial-CTE equivalent of Figure~21 in \cite{anderson2018}.) Each row of panels shows a particular background level in the short exposures, and each column of panels shows a particular bin in instrumental magnitude from the ``truth'' catalog. Background level increases from top to bottom along the rows and stellar brightness decreases from left to right along the columns. Magnitude differences between uncorrected (corrected) short exposures and ``truth'' are plotted as blue circles (orange triangles). The total number of stars in each panel is shown in the lower right corner. 

As expected for serial CTE losses, stars in the uncorrected short exposures generally lose charge (increase in magnitude) relative to ``truth'' as they experience more $x$ transfers. This is true across stellar brightness and background level, though there is more scatter in the $-7$ and $-8$~mag panels. Stars in the corrected short exposures have a shallower increase in magnitude difference as a function of $x$ transfers as compared to the uncorrected data, indicating the serial CTE correction is restoring signal to the stars. Based on the gap between the corrected and uncorrected curves at maximum $x$ transfers, the serial CTE correction restores at most $\sim$0.005 to 0.015~mag, with some dependence on stellar brightness and background level.

Two minor issues are notable in Figure~\ref{serial_phot_recovery}. For the three leftmost columns of panels ($-8$ through $-10$~mag bins), the curves are offset from zero magnitude difference by $\sim$0.01 to 0.02~mag. This offset is such that the stars in the short exposures are dimmer than expected regardless of $x$ position. \cite{ryan2024} noted a similar systematic bias ($\sim$0.02~mag) in photometry from short exposures ($t\lesssim$200s) as compared to longer exposures, as is the case here. The reason for this offset is unclear. Additionally, in the $-9$ and $-10$~mag panels, the corrected data appear to show residual signal loss as a function of $x$ transfers, i.e., the orange curves are not flat. This may indicate that the serial CTE correction has difficulty fully correcting bright stars, though it is unclear why this would be the case.

\begin{figure}[th]
  \centering
  \includegraphics[width=0.9\textwidth]{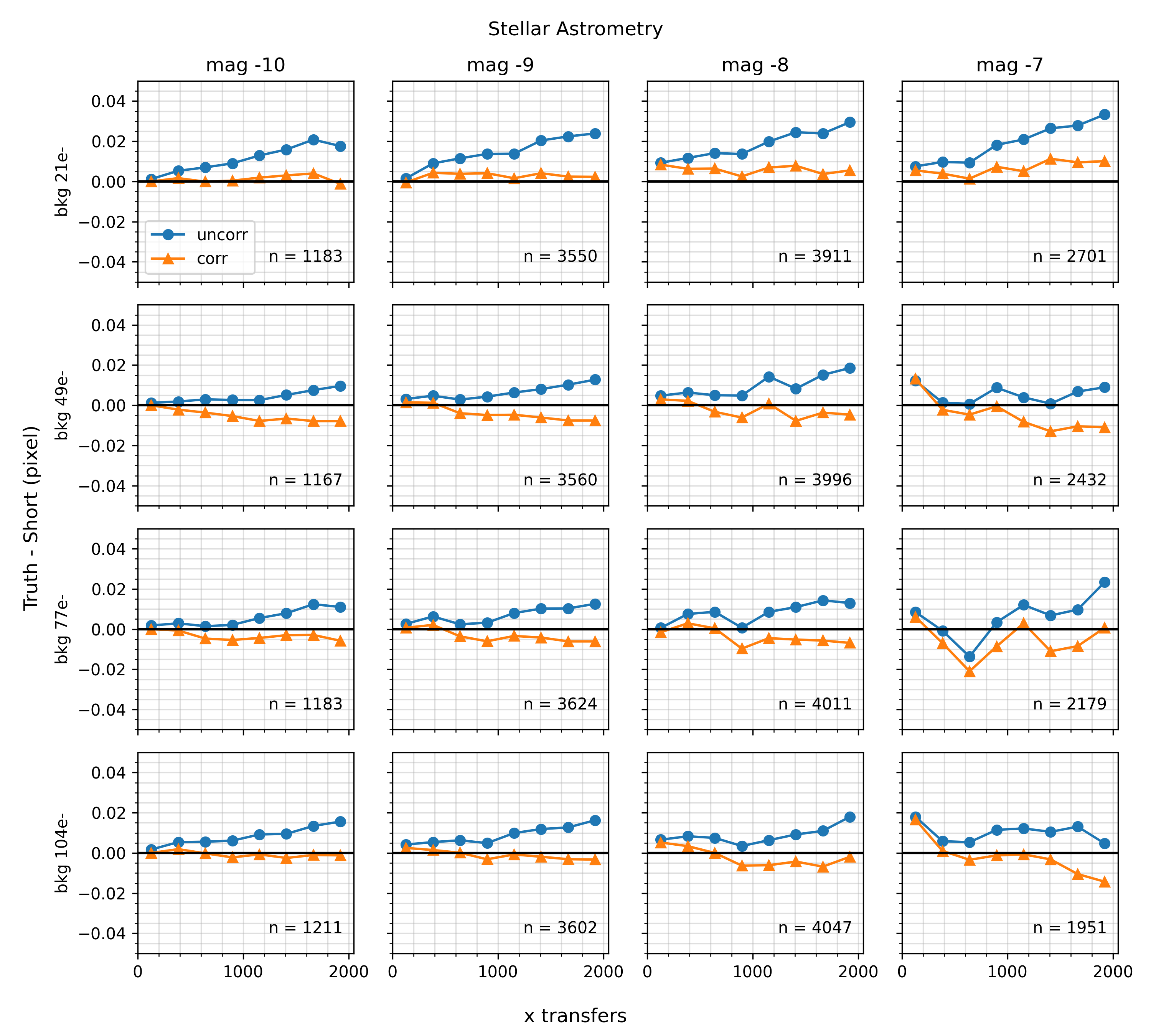}
    \caption{Same as Figure~\ref{serial_phot_recovery}, but for the difference in $x$ positions of stars between ``truth'' and short exposures.}
    \label{serial_astrometry}
\end{figure}

We compare stellar astrometry in the short and ``truth'' exposures in Figure~\ref{serial_astrometry}. Small, overall pointing offsets between the short and long exposures, ranging between $\sim$0.002 to 0.045 pixels in the $x$ direction, were corrected prior to this analysis. The difference in $x$ position of stars is plotted as a function of $x$ transfers for uncorrected (blue circles) and corrected (orange triangles) short exposures. As expected for serial CTE losses, the PSF centroids are shifted towards smaller $x$ in quadrant B subarray data as the stars experience more $x$ transfers. The effect is generally more pronounced for lower backgrounds and dimmer magnitude bins, though several of the $-7$~mag panels show significant scatter. The corrected short exposures generally show little to no centroid shift as compared to the truth data, or a slight overcorrection, suggesting that the serial CTE correction is restoring the stellar positions well. The astrometric corrections at maximum $x$ transfers are $\sim$0.018 to 0.022~pixels, depending on stellar brightness and background level.

Finally, we compare the photometry and astrometry of stars in the uncorrected and corrected exposure from CAL-17336, ignoring the idea of a ``truth'' image, to directly assess the impact of the pixel-based serial CTE correction on the data. Because this exposure is a full-frame image, we can also assess the differences between detector quadrants. We impose the same quality of fit and sharpness cuts as above on the uncorrected catalog, then match the result to the $x$ and $y$ positions of the corrected catalog.

\begin{figure}[t]
  \centering
  \includegraphics[width=0.7\textwidth]{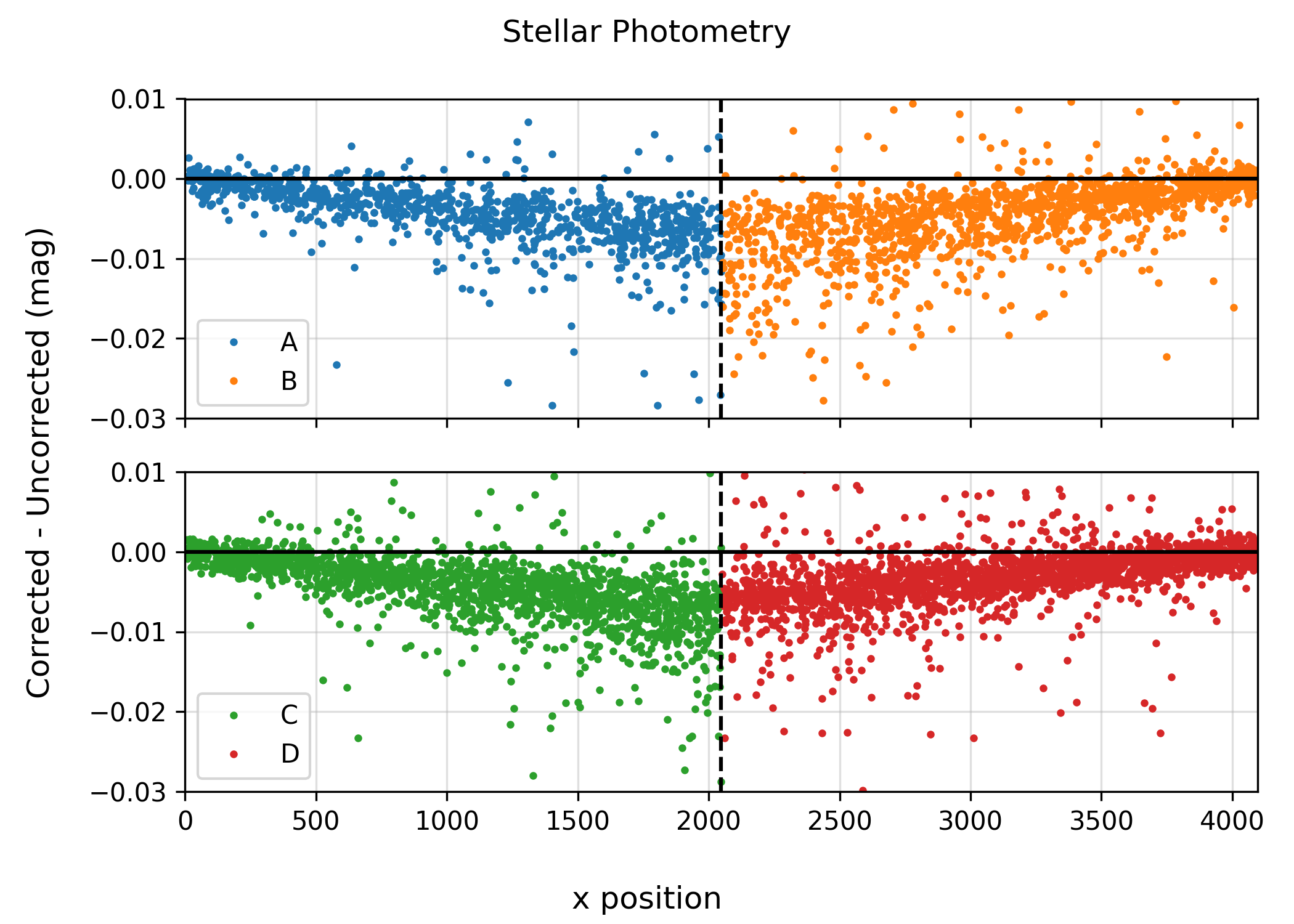}
    \caption{Magnitude differences between corrected and uncorrected stars as a function of $x$ position (ignoring ``truth''). The WFC1 quadrants A (blue) and B (orange) are plotted in the top panel, and the WFC2 quadrants C (green) and D (red) are plotted in the bottom panel. Differences worsen towards the amp gap, plotted as a dashed vertical line, as the number of $x$ transfers increases. Zero difference is shown as a solid black line.}
    \label{direct_phot}
\end{figure}

In Figure~\ref{direct_phot}, we plot the magnitude differences between the corrected and uncorrected images as a function of $x$ position. Each quadrant is represented by different color points. The differences in stellar magnitude worsen towards the amplifier gap ($x=2048$), as expected, with the typical photometric correction due to serial CTE at the amplifier gap being $\sim-0.005$ to $-0.012$~mag. Quadrant B appears to have a slightly larger correction, up to nearly $-0.02$~mag.

\begin{figure}[t]
  \centering
  \includegraphics[width=0.7\textwidth]{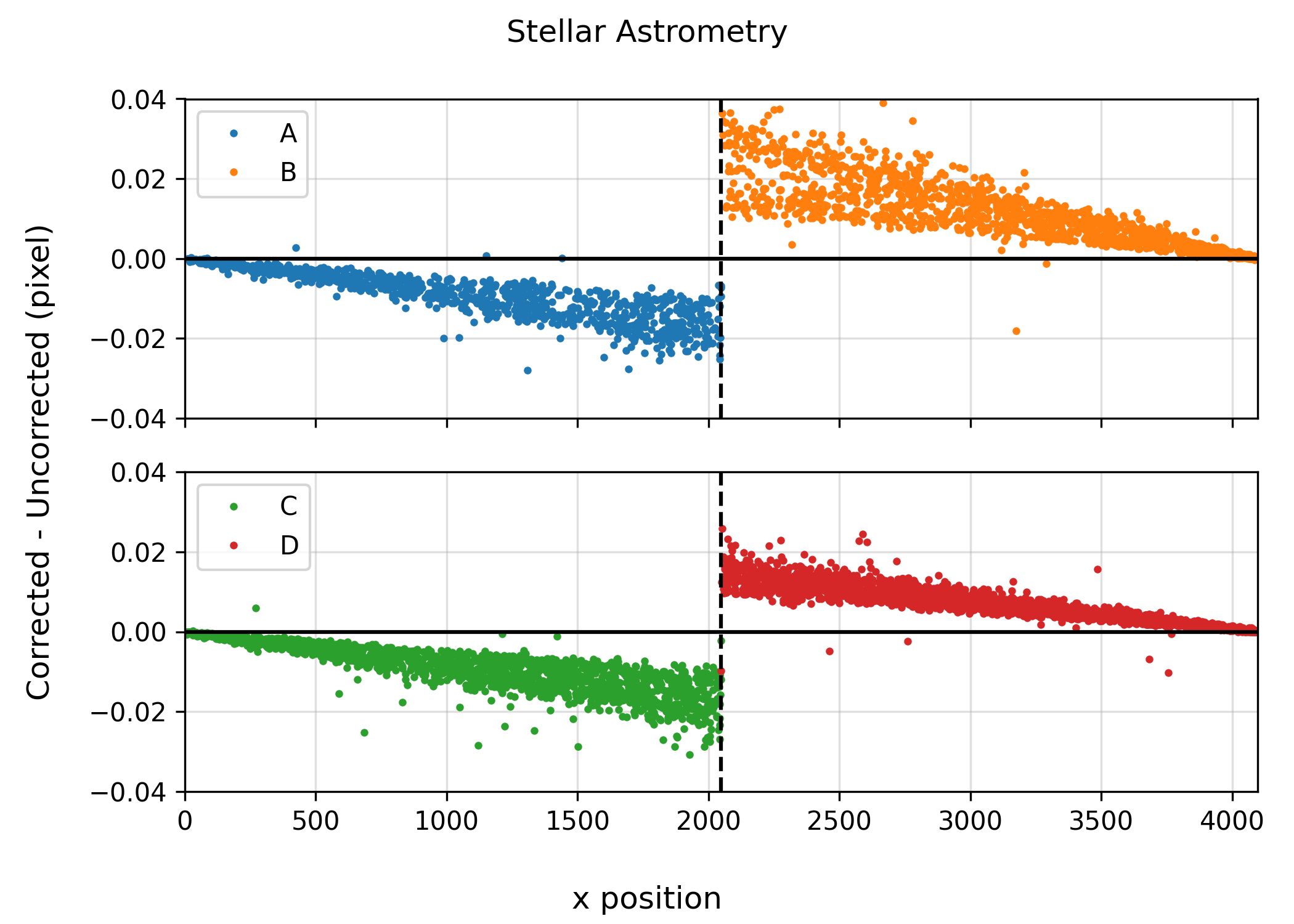}
    \caption{Same as Figure~\ref{direct_phot} for differences in $x$ position.}
    \label{direct_astrometry}
\end{figure}

In Figure~\ref{direct_astrometry}, we plot the difference in $x$ position between the corrected and uncorrected images as a function of $x$ position. The $x$ position differences are negative for quadrants A and C and positive for quadrants B and D because the serial readout is towards the left for A and C, and towards the right for B and D. The serial CTE correction therefore shifts PSF centroids towards smaller $x$ for A and C, and towards larger $x$ for B and D. As expected, the $x$ position differences increase towards the amplifier gap ($x=2048$), with the typical astrometric correction at the amplifier gap being $\sim$0.01 to 0.02 pixels (in an absolute sense). Quadrant B's astrometric correction is larger, $\sim$0.015 to 0.035 pixels.

While the effects of serial CTE losses on stellar photometry and astrometry are much smaller than that of parallel CTE losses, they may not be negligible for users who require high precision measurements. \cite{anderson2024} studied serial CTE in WFC3/UVIS and found that typical corrections for stars in UVIS data are 0.0075-0.02~pixel and 0.003-0.007~mag. Differences in serial CTE losses between the four quadrants of the UVIS detector were also found, like we find in WFC. WFC3 has been in the harsh radiation environment of space for 2/3 as long as ACS, which likely explains the smaller corrections for UVIS, despite UVIS experiencing faster degradation of parallel CTE since installation than WFC \citep{anderson2021}. 

\section{Implementation in CALACS} \label{implementation}

The pixel-based serial CTE correction algorithm is included in CALACS version 10.4.0. This version of CALACS requires the latest PCTETAB reference file, \texttt{8ch1518tj\_cte.fits}. This latest PCTETAB contains 16 additional extensions (four per amplifier) as compared to the previous version of the file, so it is incompatible with older versions of CALACS. Older versions of the PCTETAB will not work with CALACS v10.4.0 or later as well.

In the standard calibration pipeline, the serial CTE correction is run by default on all full-frame WFC exposures obtained after SM4 (May 2009). The parallel correction alone is run on pre-SM4 data, since serial CTE losses are very minor during that time period. 
The serial CTE correction is not currently available for subarray data, but we are exploring options for including it in \texttt{acs\_destripe\_plus} within \href{https://acstools.readthedocs.io/en/latest/}{\texttt{acstools}} along with the parallel CTE correction. See \href{https://hst-docs.stsci.edu/acsihb/chapter-7-observing-techniques/7-3-operating-modes#id-7.3OperatingModes-table7.6}{Table 7.6 in the ACS Instrument Handbook} for guidance on the appropriate use of \texttt{acs\_destripe\_plus} with subarray data.

If CALACS v10.4.0 (or later) is run on post-SM4 full-frame ACS/WFC data, both the serial and parallel CTE corrections will be performed. It is not possible to run one of the corrections and not the other, i.e., the PCTECORR header keyword switch to run ACSCTE controls \textbf{both} the serial and parallel corrections.

\section{Conclusions} \label{conclusions}

In this study, we characterize serial CTE in ACS/WFC, following a similar methodology to that presented in \cite{anderson2018}. The large numbers of hot pixels in calibration dark frames act as probes of charge trapping and release in the serial registers during readout. A triple exponential fit describes the serial CTE charge trails well, and provides an estimate of the probability of release of charge as a function of distance from a hot stimulus pixel for each quadrant of the detector. The fits also provide an estimate of the number of charge traps encountered by charge packets of various sizes for each quadrant of the detector. We manually adjust these two parameter sets and the time-dependence of the model until the serial-CTE-corrected dark frames show little to no residual serial CTE trailing from hot pixels. 

The pixel-based algorithm in CALACS that corrects for parallel CTE losses in ACS/WFC data has been modified to include a correction for serial CTE losses as well. The PCTETAB reference file has been updated to include the tabulated parameters for serial CTE correction. Both are installed in the standard calibration pipeline and the serial CTE correction now runs by default on full-frame post-SM4 (after 2009) ACS/WFC data. Shortly following the publication of this report, science data corrected for both parallel and serial CTE will be available in the MAST archive.

We find that stellar photometry and astrometry are improved by serial CTE correction. In recent data, photometric corrections are about 0.005-0.02 magnitudes, and astrometric corrections are 0.01-0.035 pixels, with some dependence on stellar brightness and local background signal. Unexplained, residual signal loss was seen in photometry of bright stars from serial-CTE-corrected data, and may be addressed in future work.

\section*{Acknowledgements}

Thanks to Michele De La Pe\~na for her time and effort implementing the serial CTE correction in CALACS. We also thank Jay Anderson and Yotam Cohen for enlightening discussions throughout this project. Thanks to the following members of the ACS team for their helpful comments on this report: David Stark, Nimish Hathi, Gagandeep Anand, and Meaghan McDonald.

This work made use of \texttt{acstools} \citep{lim2020}, \texttt{jupyter} \citep{kluyver2016}, \texttt{numpy} and \texttt{scipy} \citep{virtanen2019}, \texttt{pandas} \citep{mckinney2010}, \texttt{astropy} \citep{astropy:2013,astropy:2018}, \texttt{matplotlib} \citep{hunter2007}, \texttt{KDEpy} \citep{odland2018}, and \texttt{dask} \citep{dask2016}.

\bibliography{serial_cte}
\bibliographystyle{apj}

\end{document}